\documentclass[prd,reprint,superscriptaddress,amsmath,nofootinbib,twocolumn]{revtex4-1}
\pdfoutput=1

\usepackage{ifthen}
\usepackage{epsfig}
\usepackage{booktabs} 
\usepackage{natbib}
\usepackage{wasysym}
\usepackage{feynmp}
\usepackage{slashed} 
\usepackage{bbm}
\usepackage{mathrsfs}
\usepackage{amssymb}
\usepackage{dsfont}
\usepackage[export]{adjustbox}
\usepackage{bm}
\usepackage{soul}

\newcommand{\boldmidrule}{\specialrule{1.5pt}{0em}{0em}}

\usepackage{color}

\newcommand{\beqra}{\begin{eqnarray}}
\newcommand{\eeqra}{\end{eqnarray}}
\newcommand{\beq}{\begin{equation}}
\newcommand{\eeq}{\end{equation}}

\renewcommand{\epsilon}{\varepsilon}

\renewcommand{\bar}{\overline}

\begin{document}

\title{$J$-factors for self-interacting dark matter in 20 dwarf spheroidal galaxies}

\author{Sebastian Bergstr\"om}
\affiliation{Chalmers University of Technology, Department of Physics, SE-412 96 G\"oteborg, Sweden}
\author{Riccardo Catena}
\affiliation{Chalmers University of Technology, Department of Physics, SE-412 96 G\"oteborg, Sweden}
\author{Andrea Chiappo}
\affiliation{Oskar Klein Centre, Department of Physics, Stockholm University, AlbaNova, Stockholm SE-10691, Sweden}
\author{Jan Conrad}
\affiliation{Oskar Klein Centre, Department of Physics, Stockholm University, AlbaNova, Stockholm SE-10691, Sweden}
\author{Bj\"orn Eurenius}
\affiliation{University of Gothenburg, Department of Physics, Origov\"agen 6 B SE-412 96 G\"oteborg, Sweden}
\affiliation{Chalmers University of Technology, Department of Physics, SE-412 96 G\"oteborg, Sweden}
\author{Magdalena Eriksson}
\affiliation{Chalmers University of Technology, Department of Physics, SE-412 96 G\"oteborg, Sweden}
\author{Michael H\"ogberg}
\affiliation{Chalmers University of Technology, Department of Physics, SE-412 96 G\"oteborg, Sweden}
\author{Susanna Larsson}
\affiliation{Chalmers University of Technology, Department of Physics, SE-412 96 G\"oteborg, Sweden}
\author{Emelie Olsson}
\affiliation{University of Gothenburg, Department of Physics, Origov\"agen 6 B SE-412 96 G\"oteborg, Sweden}
\affiliation{Lund University, Department of Physics, Box 118, 221 00 Lund, Sweden}
\author{Andreas Unger}
\affiliation{Chalmers University of Technology, Department of Physics, SE-412 96 G\"oteborg, Sweden}

\author{Rikard Wadman}
\email{sebbergs@student.chalmers.se\\catena@chalmers.se\\guseurbj@student.gu.se\\
andrea.chiappo@fysik.su.se\\conrad@fysik.su.se\\
magderi@student.chalmers.se\\michog@student.chalmers.se\\suslars@student.chalmers.se\\gusolsemv@student.gu.se\\ungera@student.chalmers.se\\rikardw@student.chalmers.se}
\affiliation{Chalmers University of Technology, Department of Physics, SE-412 96 G\"oteborg, Sweden}

\begin{abstract} 
Dwarf spheroidal galaxies are among the most promising targets for indirect dark matter (DM) searches in $\gamma$-rays.~The $\gamma$-ray flux from DM annihilation in a dwarf spheroidal galaxy is proportional to the $J$-factor of the source.~The $J$-factor of a dwarf spheroidal galaxy is the line-of-sight integral of the DM mass density squared times $\langle \sigma_{\rm ann} v_{\rm rel} \rangle/(\sigma_{\rm ann} v_{\rm rel})_0$, where $\sigma_{\rm ann} v_{\rm rel}$ is the DM annihilation cross-section times relative velocity $v_{\rm rel}=|\bm{v}_{\rm rel}|$, angle brackets denote average over $\bm{v}_{\rm rel}$, and $(\sigma_{\rm ann} v_{\rm rel})_0$ is the $v_{\rm rel}$-independent part of $\sigma_{\rm ann} v_{\rm rel}$.~If $\sigma_{\rm ann} v_{\rm rel}$ is constant in $v_{\rm rel}$, $J$-factors only depend on the DM space distribution in the source.~However, if $\sigma_{\rm ann} v_{\rm rel}$ varies with $v_{\rm rel}$, as in the presence of DM self-interactions, $J$-factors also depend on the DM velocity distribution, and on the strength and range of the DM self-interaction.~Models for self-interacting DM are increasingly important in the study of the small scale clustering of DM, and are compatible with current astronomical and cosmological observations.~Here we derive the $J$-factor of 20 dwarf spheroidal galaxies from stellar kinematic data under the assumption of Yukawa DM self-interactions.~$J$-factors are derived through a profile Likelihood approach, assuming either NFW or cored DM profiles.~We also compare our results with $J$-factors derived assuming the same velocity for all DM particles in the target galaxy.~We find that this common approximation overestimates the $J$-factors by up to one order of magnitude.~$J$-factors for a sample of DM particle masses and self-interaction coupling constants, as well as for NFW and cored density profiles, are provided electronically, ready to be used in other projects.
\end{abstract}

\maketitle

\section{Introduction}
\label{sec:introduction}
Increasingly accurate cosmological and astronomical data indicate that the Universe is to a large extent made of a non-luminous component called dark matter (DM)~\cite{Bertone:2016nfn}.~The nature of DM remains a mystery, but indirect evidence points towards a new hypothetical particle as the prime candidate~\cite{Bertone:2004pz}.~If DM is made of Weakly Interacting Massive Particles (WIMPs) -- the leading paradigm in modern cosmology -- it can pair annihilate into Standard Model particles which can in turn be observed with satellites or ground-based detectors~\cite{Roszkowski:2017nbc,Arcadi:2017kky}.~This is the essence of an experimental technique known as DM indirect detection~\cite{Gaskins:2016cha}.~Different annihilation products are currently searched for experimentally, including electron-positron pairs, quark-antiquark pairs, and pairs of gauge bosons, such as $\gamma$-ray photons.~The $\gamma$-ray channel benefits from the geodesic motion of photons, which do not diffuse in the galactic magnetic field, in contrast to charged annihilation products~\cite{Bringmann:2012ez}.~Dwarf spheroidal galaxies (dSphs) are among the most promising targets for indirect DM searches in $\gamma$-rays~\cite{Lefranc:2016fgn}.~Indeed, dSphs are DM dominated astrophysical objects, a property implying a large signal to noise ratio~\cite{Conrad:2015bsa}. 

The flux of $\gamma$-ray photons from DM annihilation in dSphs is proportional to the line-of-sight integral of the annihilation rate $\Gamma_\gamma \propto \rho_\chi^2 \langle \sigma_{\rm ann} v_{\rm rel} \rangle$~\cite{Bringmann:2012ez}, where $\rho_\chi$ is the DM density along the line of sight, $\sigma_{\rm ann}$ is the DM annihilation cross-section, $v_{\rm rel}=|\bm{v}_{\rm rel}|$ is the DM-DM relative speed, and angle brackets denote an average over the three-dimensional DM-DM relative velocity distribution $P_{r,\rm rel}(\bm{v}_{\rm rel})$ at a distance $r$ from the centre of the dSph.~If $\sigma_{\rm ann} v_{\rm rel}$ is independent of $v_{\rm rel}$, as for $S$-wave DM annihilations~\cite{Drees:1992am}, $\rho_\chi^2$ is the only term in $\Gamma_\gamma$ which depends on the line-of-sight coordinate $s$, since $P_{r,\rm rel}$ trivially disappears integrating over $\bm{v}_{\rm rel}$, e.g.~\cite{Ferrer:2013cla}.~This simplification leads to the canonical definition of $J$-factor:~the $J$-factor of a dSph is the integral of $\rho_\chi^2$ along the line-of-sight and over the angular size of the target galaxy.~$J$-factors are crucial in DM indirect detection, since the flux of $\gamma$-ray photons from DM annihilation in dSphs is proportional to $J$.~Assuming $\sigma_{\rm ann} v_{\rm rel}$ is independent of $v_{\rm rel}$, $J$-factors have been computed for relatively large samples of dSphs, e.g.~\cite{Geringer-Sameth:2014yza,Bonnivard:2014kza,Bonnivard:2015pia}.~One of the key aspects in this calculation is the error propagation:~from the velocity of individual stars tracing the total gravitational potential and DM distribution in the dSph to the $J$-factor~\cite{Ullio:2016kvy}.~In this context, a profile likelihood approach has recently been proposed in Ref.~\cite{Chiappo:2016xfs}.~The advantage of this approach is that it does not depend on priors, unlike previous Bayesian analyses.

On the other hand, there exist many well-motivated particles physics models where $\sigma_{\rm ann} v_{\rm rel}$ varies with $v_{\rm rel}$, and for which the flux of $\gamma$-ray photons from DM annihilation in dSphs explicitly depends on $P_{r,\rm rel}$.~Examples include models where DM primarily annihilate via $P$-wave~\cite{Drees:1992am} or resonant processes~\cite{Ibe:2008ye}, or where DM self-interacts, e.g.~\cite{Carlson:1992fn,McDonald:2001vt,Ackerman:mha,Buckley:2009in,Bellazzini:2013foa,Cyr-Racine:2015ihg}.~In the latter case, $\langle \sigma_{\rm ann} v_{\rm rel} \rangle$ is a non trivial function of $P_{r,\rm rel}({\bm v}_{\rm rel})$~\cite{Robertson:2009bh,Ferrer:2013cla}.~Consequently, $\rho_\chi^2$ is not the only term depending on the line-of-sight coordinate $s$ in the annihilation rate $\Gamma_\gamma$.~Accordingly, the canonical definition of $J$-factor given above needs to be generalised.~In the case of, e.g., self-interacting DM, the generalised $J$-factor of dSphs is the integral along the line of sight, over the angular coordinates $(\theta,\phi)$ of the target galaxy, and over $\bm{v}_{\rm rel}$ of $P_{r,\rm rel} (\bm{v}_{\rm rel}) S(v_{\rm rel})\rho_\chi^2(s,\theta)$, where the radial coordinate $r(s)$ is a function of $s$, and $S(v_{\rm rel})$ is a model-dependent particle physics input, e.g.~the Sommerfeld enhancement in the case of DM self-interactions~\cite{ArkaniHamed:2008qn}.

The aim of this work is to derive the generalised $J$-factor, $J_S$, of 20 dSphs from stellar kinematic data in the case of self-interacting DM ($S(v_{\rm rel})\neq 1$)~\cite{Carlson:1992fn,McDonald:2001vt,Ackerman:mha,Buckley:2009in,Bellazzini:2013foa,Cyr-Racine:2015ihg}.~Specifically, we consider a family of DM self-interactions which in the non-relativistic limit is described by a Yukawa potential~\cite{Loeb:2010gj}.
DM-self interactions have recently been considered as one of the possible solutions to the $\Lambda$CDM ``small scale crisis'', e.g.,~\cite{Vogelsberger:2015gpr,Loeb:2010gj} -- the mismatch between observations on the scale of dwarf galaxies (or below) and predictions for the clustering of DM based on DM only N-body simulations.~Although alternative explanations exist, e.g., feedback from supernovae explosion in hydrodynamical simulations~\cite{Read:2015sta,Pontzen:2011ty}, DM self-interaction cross-sections per unit DM particle mass of the order of $10^{-24}$~cm$^2$/GeV remain compatible with astronomical and cosmological observations~\cite{Mahoney:2017jqk,Kahlhoefer:2017umn,Bringmann:2016din,Kahlhoefer:2015vua,Harvey:2015hha}, and deserve further exploration.

We determine $J_S$ and the associated statistical error within the profile likelihood approach proposed in~\cite{Chiappo:2016xfs,James:2006zz}.~Results are presented for NFW and cored DM profiles and for different combinations of particle physics parameters.~We find significant differences between canonical and generalised $J$-factors -- up to several orders of magnitude for all dSphs.~We also compare our results with the common approximation made when calculating $\gamma$-ray fluxes from DM annihilation in dSphs: $S(v_{\rm rel})=S(v^*)$, where $v^*$ is a reference velocity for DM particles in dSphs.~This approximation corresponds to assigning to all DM particles in a dSph the same reference velocity.~We find that this approximation leads to overestimate $J_S$, with errors as large as one order of magnitude.

So far, only Ref.~\cite{Boddy:2017vpe} has used stellar kinematic data to obtain $J_S$.~Ref.~\cite{Boddy:2017vpe} focuses of 4 dSphs, computes $S(v_{\rm rel})$ within the Hulth\'en approximation, presents results for NFW profiles, and does not rely on a profile likelihood approach.~Here we extend the analysis in~\cite{Boddy:2017vpe} to 20 galaxies, computing $S(v_{\rm rel})$ numerically, presenting results for NFW and cored DM profiles, and deriving $J_S$ and associated statistical error through a profile likelihood approach.

This work is organised as follows.~In Sec.~\ref{sec:gamma} we introduce the generalised $J$-factor, $J_S$.~In Sec.~\ref{sec:analysis} we describe our method to determine $J_S$ from stellar kinematic data.~Results are presented and discussed in Sec.~\ref{sec:results}.~We conclude in Sec.~\ref{sec:conclusions}.

\section{$\gamma$-rays from the annihilation of self-interacting DM}
\label{sec:gamma}

\subsection{$\gamma$-ray flux}
The $\gamma$-ray flux from DM annihilation in dSphs can be written as follows\footnote{If the DM particle and anti-particle are distinct, as for Dirac fermions, and equally abundant, Eq.~(\ref{eq:flux}) must include an additional factor 1/2.}
\begin{equation}
\frac{{\rm d} \Phi_\gamma}{{\rm d}E_\gamma} = \frac{1}{8\pi} \frac{{\rm d} N}{{\rm d} E_\gamma} \int_{\Delta \Omega} {\rm d}\Omega\int_{\rm l.o.s.} {\rm d}s \int {\rm d}^3 \bm{v}_{\rm rel} \, \mathscr{J}(s,\theta,\bm{v}_{\rm rel}) \,, \label{eq:flux}
\end{equation}
where 
\begin{equation}
\mathscr{J}(s,\theta,\bm{v}_{\rm rel}) = n_\chi^2(s,\theta) \,P_{r(s,\theta),\rm rel} (\bm{v}_{\rm rel})  \, \sigma_{\rm ann} v_{\rm rel} \,.
\label{eq:jcal}
\end{equation}
In Eq.~(\ref{eq:flux}), $\rm d\Omega=\sin\hspace{-0.05cm}\theta{\rm d}\theta {\rm d}\phi$, where $\theta\in[0,\theta_{\rm max}]$ and $\phi\in[0,2\pi]$ are, respectively, the polar and azimuthal angle of a spherical coordinate system with z-axis along the line-of-sight, $2\theta_{\rm max}$ is the angular diameter of the dSph, and $s$ is the line-of-sight coordinate.~As already anticipated, $P_{r(s,\theta),\rm rel} (\bm{v}_{\rm rel})$ is the three-dimensional DM-DM relative velocity distribution at the radial distance from the dSph $r(s,\theta)=\sqrt{D^2+s^2-2 D s\cos\theta}$, $v_{\rm rel}=|\bm{v}_{\rm rel}|$, and $D$ is the distance from the observer to the centre of the system.~In Eq.~(\ref{eq:jcal}), $n_\chi=\rho_{\chi}/m_\chi$ is the DM number density, and $m_\chi$ is the DM particle mass.~Finally, $\sigma_{\rm ann}$ is the DM annihilation cross-section, ${\rm d} N/{\rm d} E_\gamma$ the differential $\gamma$-ray photon yield per DM pair annihilation, and $E_\gamma$ the photon energy.

\subsection{Annihilation cross-section and Sommerfeld enhancement}
\label{sec:theory}
We calculate the cross-section $\sigma_{\rm ann}$ assuming that DM can pair annihilate into $\gamma$-ray photons, like ordinary WIMPs.~We also assume that DM is self-interacting~\cite{Carlson:1992fn,McDonald:2001vt,Ackerman:mha,Buckley:2009in,Bellazzini:2013foa,Cyr-Racine:2015ihg} and that in the non-relativistic limit DM self-interactions are characterised by the attractive Yukawa potential
\begin{equation}
\mathscr{V}(\rho) = - \frac{\alpha_\chi}{\rho} e^{-m_\phi \rho}\,,
\label{eq:V}
\end{equation}
where $\rho$ is the relative distance between two annihilating DM particles, $\alpha_\chi$ is a positive coupling constant and $m_{\phi}$ is the mass of the particle that mediates the DM self-interaction. 

Let us denote by $(\sigma_{\rm ann})_0$ the DM annihilation cross-section in the limit $\alpha_\chi=0$, when self-interactions are negligible.~For $\alpha_\chi\neq0$, $\sigma_{\rm ann}\neq(\sigma_{\rm ann})_0$, since the wave function $\psi_k$ describing the relative motion of the annihilating DM particles is perturbed by the Yukawa interaction in Eq.~(\ref{eq:V}).~Since the cross-section $\sigma_{\rm ann}$ depends on $\psi_k$ quadratically~\cite{ArkaniHamed:2008qn}, 
\begin{equation}
\sigma_{\rm ann} = S(v_{\rm rel}) (\sigma_{\rm ann})_0\,,
\end{equation} 
where
\begin{equation}
S(v_{\rm rel}) =|\psi_k|^2_{\rho=0} \,,
\end{equation}
and $k=m_\chi v_{\rm rel}/2$.~The velocity dependent factor $S(v_{\rm rel})$ is also known as Sommerfeld enhancement~\cite{ArkaniHamed:2008qn}.~We calculate $S(v_{\rm rel})$ by numerically solving the radial Schr\"odinger equation
\begin{equation} 
\left[ - \frac{1}{\rho} \frac{{\rm d}^2}{{\rm d} \rho^2} \rho + \frac{\ell (\ell+1)}{\rho^2}  - k^2 +m_\chi\mathscr{V}(\rho) \right] R_{k\ell}(\rho) = 0 \,,
\label{eq:SE}
\end{equation}
where $R_{k\ell}(\rho)$ denotes the radial part of the wave function $\psi_k$,
\begin{equation}
\psi_k(\rho,\Theta)= \sum_{\ell=0}^{\infty} i^\ell (2\ell+1) e^{i\delta_\ell} R_{k\ell}(\rho)P_{\ell}(\cos\Theta)\,,
\label{eq:series}
\end{equation}
$\Theta$ is the polar angle of a spherical coordinate system with $z$-axis in the direction of the relative motion, and $\delta_\ell$ are phase shifts.~We solve Eq.~(\ref{eq:SE}) imposing the boundary conditions 
\begin{eqnarray}
&&\lim_{\rho\rightarrow 0} k\rho \, R_{k\ell}(\rho) = 0\,,\\
&&\lim_{\rho\rightarrow \infty} \frac{k\rho\, R_{k\ell}(\rho)}{\mathcal{C}_\ell\sin\left(k\rho -\frac{1}{2}\pi \ell+ \delta_\ell\right)} = 1\,.
\label{eq:bound}
\end{eqnarray}
In Eq.~(\ref{eq:bound}), $\mathcal{C}_\ell$ is a normalisation constant.~In what follows, we focus on the case $\ell=0$, i.e.~S-wave Sommerfeld enhanced DM annihilation, so that $S=|1/\mathcal{C}_0|^2$~\cite{ArkaniHamed:2008qn}.~Notice that in the numerical calculations it is convenient to introduce the new variable $\chi(x)=k\rho\,R_{k0}(\rho)$, with $x=\alpha_\chi m_\chi \rho$.~The function $\chi$ obeys the one-dimensional equation
\begin{equation} 
\frac{{\rm d}^2}{{\rm d}x^2}\chi(x) + \left[ \varepsilon_{v}^2 +\mathscr{U}(x) \right] \chi(x) = 0 \,,
\label{eq:SE2}
\end{equation}
where
\begin{equation}
\mathscr{U}(x) = - \frac{1}{x} e^{-\varepsilon_\phi x}\,.
\label{eq:U}
\end{equation}
Solutions to Eq.~(\ref{eq:SE2}) only depend on the dimension-less parameters
\begin{eqnarray}
&&\varepsilon_v=\frac{v_{\rm rel}}{2\alpha_\chi} \,,\\
&&\varepsilon_\phi=\frac{m_\phi}{\alpha_\chi m_\chi} \,.
\end{eqnarray}
For further details on the numerical solution of Eq.~(\ref{eq:SE2}), we refer to~\cite{Iengo:2009ni,Iengo:2009xf}.~The Sommerfeld enhancement $S$ depends on $\epsilon_v$ and $\epsilon_\phi$ as follows:~a) For $\epsilon_\phi\ll \epsilon_v$, $S\simeq \pi/\epsilon_v= 2 \pi \alpha_\chi/v_{\rm rel}$ (Coulomb limit); b) For $\epsilon_v\ll \epsilon_\phi$ and $\epsilon_\phi <1$, $S\simeq 12/\epsilon_\phi = 12 \alpha_\chi m_X/m_\phi$, unless $\epsilon_\phi \simeq 6/(\pi^2 n^2)$, $n\in \mathbb{Z}^+$, in which case $S\simeq 4 \alpha_\chi^2/(v_{\rm rel}^2 n^2)$; c) Finally, for $\epsilon_\phi \gg 1$ there is no Sommerfeld enhancement and consequently $S=1$.

\subsection{Dark matter velocity distribution}
\label{sec:DF}
The velocity distribution $\mathscr{P}_{r,\rm rel} (\bm{v}_{\rm rel})$ in Eq.~(\ref{eq:flux}) can be written as
\begin{equation}
P_{r,\rm rel} (\bm{v}_{\rm rel})=\int {\rm d}^3\bm{v}_{\rm cm} \, \mathscr{P}_{r,{\rm pair}}(\bm{v}_{\rm cm}, \bm{v}_{\rm rel})
\label{eq:Prel}
\end{equation}
where
\begin{align}
\mathscr{P}_{r,{\rm pair}}(\bm{v}_{\rm cm}, \bm{v}_{\rm rel}) &= \mathscr{P}_{r}(\bm{v}_{1}) \mathscr{P}_{r}(\bm{v}_{2}) \nonumber\\
&= \mathscr{P}_{r}(\bm{v}_{\rm cm} + \bm{v}_{\rm rel}/2) \mathscr{P}_{r}(\bm{v}_{\rm cm} - \bm{v}_{\rm rel}/2)\,. 
\label{eq:Ppair}
\end{align}
In the above equations, $\mathscr{P}_{r}$ is the DM single particle velocity distribution at $r$, and $\bm{v}_{\rm cm}=(\bm{v}_1 + \bm{v}_2)/2$ and $\bm{v}_{\rm rel}=(\bm{v}_1 - \bm{v}_2)$ are the centre-of-mass and relative velocities, respectively.~Accordingly, $\bm{v}_1=\bm{v}_{\rm cm} + \bm{v}_{\rm rel}/2$ and $\bm{v}_2=\bm{v}_{\rm cm} - \bm{v}_{\rm rel}/2$.~Using spherical coordinates with z-axis along the direction of $\bm{v}_{\rm rel}$, and assuming isotropy for $\mathscr{P}_r$, i.e.~$\mathscr{P}_r(\bm{v}_1)=\mathscr{P}_r(|\bm{v}_1|)$, Eq.~(\ref{eq:Prel}) can be expressed as follows
\begin{equation}
\mathscr{P}_{r,{\rm rel}}(\bm{v}_{\rm rel}) = 2\pi \int_0^{\infty} {\rm d} v_{\rm cm} v_{\rm cm}^2 \int_{-1}^{+1} {\rm d}z\, \mathscr{P}_{r}(V_{z^+})\mathscr{P}_{r}(V_{z^-})
\end{equation}
where $|\bm{v}_1|=V_{z^{+}}$, $|\bm{v}_2|=V_{z^{-}}$,
\begin{equation}
V_{z^{\pm}}^2 = \left( v_{\rm cm}^2 +\frac{v_{\rm rel}^2}{4} \pm v_{\rm cm} v_{\rm rel} z \right) \,,
\end{equation}
and $z=\bm{v}_{\rm cm} \cdot \bm{v}_{\rm rel}/(v_{\rm cm} v_{\rm rel})$, with $v_{\rm cm}=|\bm{v}_{\rm cm}|$ and $v_{\rm rel}=|\bm{v}_{\rm rel}|$.~The $v_{\rm rel}$ distribution, 
$\mathscr{P}_{r,{\rm rel}}(v_{\rm rel})$, is then simply given by
\begin{equation}
\mathscr{P}_{r,{\rm rel}}(v_{\rm rel}) = 4\pi v_{\rm rel}^2 \mathscr{P}_{r,{\rm rel}}(\bm{v}_{\rm rel}) \,,
\end{equation}
which follows from the isotropy of the single particle velocity distribution, $\mathscr{P}_{r}$.

Assuming spherical symmetry for $\rho_\chi$ in addition to isotropy in the single particle velocity space, $\mathscr{P}_{r}$ can be expressed as follows
\begin{equation}
\mathscr{P}_{r}(v) = \frac{1}{\sqrt{8}\pi^2 \rho_\chi(r)} \int_{\Psi^{-1}(\mathcal{E}(r,v))}^{\infty} \frac{{\rm d}\bar{r}}{\sqrt{\mathcal{E}(r,v)-\Psi(\bar{r})}} \,\mathscr{F}(\bar{r}) \,,
\label{eq:F}
\end{equation}
where $v=|\bm{v}_1|$ (or $v=|\bm{v}_2|$) and
\begin{equation}
\mathscr{F}(r)= \left[ \frac{{\rm d} \rho_\chi}{{\rm d} r}\frac{{\rm d}^2 \Psi}{{\rm d} r^2}\left(\frac{{\rm d}\Psi}{{\rm d}r}\right)^{-2} -\frac{{\rm d}^2\rho_\chi}{{\rm d}r^2}\left(\frac{{\rm d}\Psi}{{\rm d}r^2} \right)^{-1}\right]\,.
\label{eq:Fint}
\end{equation}
In Eq.~(\ref{eq:Fint})$, \Psi(r)=\Phi(\infty)-\Phi(r)$, while $\Phi(r)$ and $\mathcal{E}(r,v)=~(1/2)v^2+\Psi(r)$ correspond to total gravitational potential and relative energy at $r$, respectively.~Here we assume that only DM contributes to $\Psi(r)$.~Interestingly, Eq.~(\ref{eq:F}) represents the unique solution to the integral equation 
\begin{equation}
4\pi \int_0^\infty {\rm d}v \,v^2 \mathscr{P}_{r}(v) = 1\,,
\end{equation}
which simultaneously solves the Vlasov equation for the DM phase-space density $F(r,v)= \rho_\chi(r) \mathscr{P}_{r}(v)$, and which is compatible with the Poisson equation linking $\rho_\chi(r)$ to $\Phi(r)$.~Eqs.~(\ref{eq:Prel}) and (\ref{eq:F}) must be modified if the distribution $\mathscr{P}_r$ is anisotropic.~We will extend the present analysis to anisotropic velocity distributions in a future work.

As far as the DM mass density is concerned, we assume the profile
\begin{equation}
\rho_\chi(r)= \rho_0 \left(\frac{r_0}{r}\right)^\gamma \left[ 1+\left(\frac{r}{r_0}\right)^\alpha \,\right]^{\frac{\gamma-\beta}{\alpha}} \,,
\label{eq:prof}
\end{equation}
and focus on the $(\alpha,\beta,\gamma)=(1,3,1)$ and $(\alpha,\beta,\gamma)=(1,3,0)$ cases, corresponding to NFW~\cite{Navarro:1995iw} and cored Zhao~\cite{Zhao:1995cp} profile, respectively.

\subsection{Definition of generalised $J-$ factor}
If DM is self-interacting (i.e.~$S(v_{\rm rel})\neq 1$), Eq.~(\ref{eq:flux}) can be written as 
\begin{equation}
\frac{{\rm d} \Phi_\gamma}{{\rm d}E_\gamma} = \frac{1}{8\pi} \frac{{\rm d} N}{{\rm d} E_\gamma} \,(\sigma_{\rm ann})_0J_S \,,
\end{equation}
where
\begin{equation}
J_S=\int_{\Delta \Omega} {\rm d}\Omega\int_{\rm l.o.s.} {\rm d}s \int {\rm d}^3 \bm{v}_{\rm rel} \, \widetilde{\mathscr{J}}(s,\theta,\bm{v}_{\rm rel})
\label{eq:JS}
\end{equation}
and 
\begin{eqnarray}
\widetilde{\mathscr{J}}(s,\theta,\bm{v}_{\rm rel}) = n_\chi^2(s,\theta) \,P_{r(s,\theta),\rm rel} (\bm{v}_{\rm rel})  \, S(v_{\rm rel}) \,. \nonumber\\
\end{eqnarray}
Explicitly, the angular integration in Eq.~(\ref{eq:JS}) is performed as follows 
\begin{equation}
\int_{\Delta \Omega} {\rm d}\Omega = 2\pi \int_{\cos\hspace{-0.05cm}\theta_{\rm max}}^1 {\rm d}\hspace{-0.05cm}\cos\hspace{-0.05cm}\theta \,,
\end{equation}
where for $\theta_{\rm max}$ we assume $\theta_{\rm max}=0.5^\circ$.~We will refer to $J_S$ as generalised $J$-factor.~With the definition in Eq.~(\ref{eq:JS}), generalised and canonical $J$-factors coincide in the $S(v_{\rm rel})\rightarrow 1$ limit, i.e.~no self-interaction.~As already anticipated in Sec.~\ref{sec:introduction}, the aim of this work is to derive the generalised $J$-factor of 20 dSphs from stellar kinematic data. 

\begin{figure*}[t]
\begin{center}
\begin{minipage}[t]{0.49\linewidth}
\centering
\includegraphics[width=0.9\textwidth]{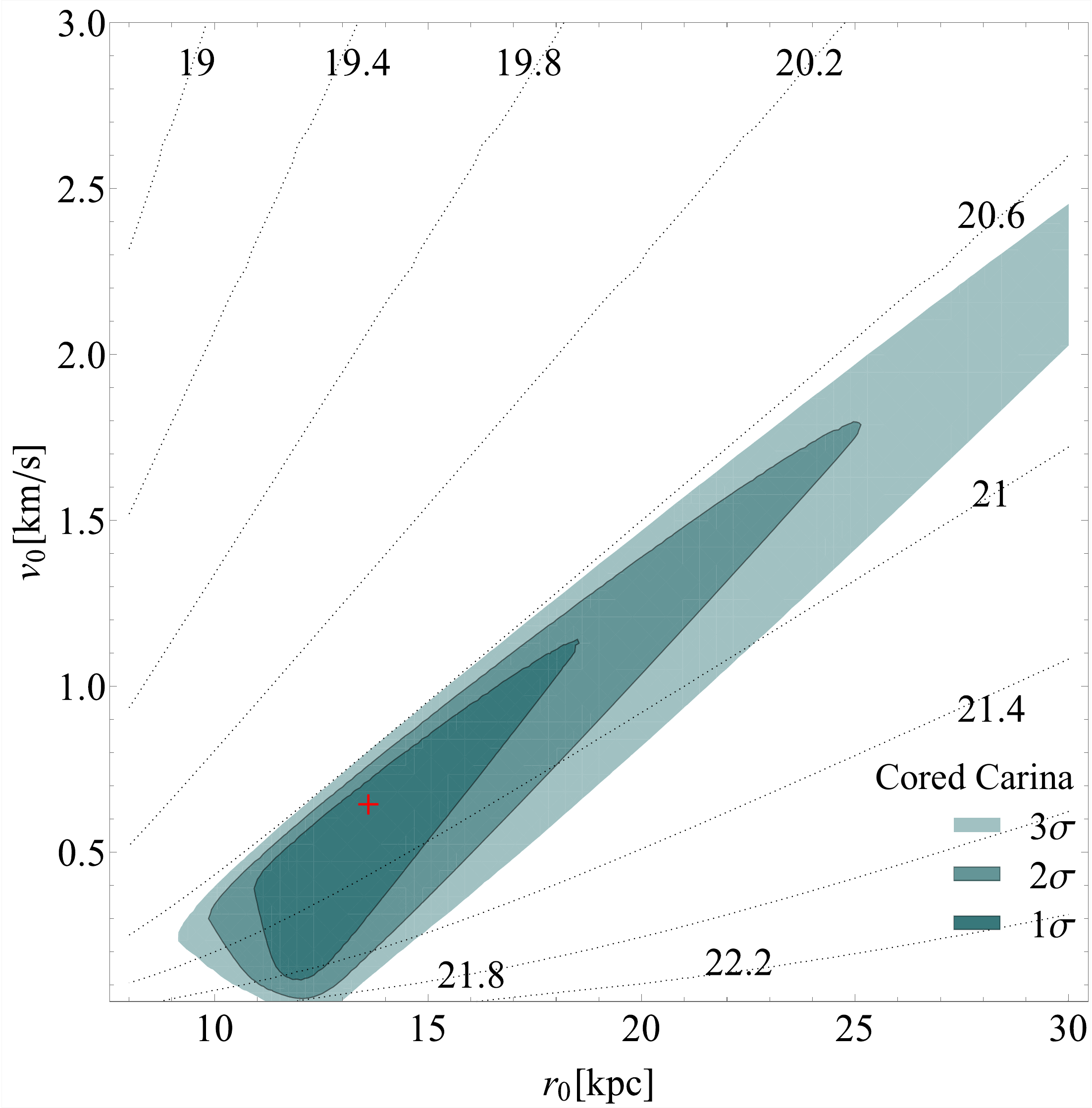}
\end{minipage}
\begin{minipage}[t]{0.49\linewidth}
\centering
\includegraphics[width=0.9\textwidth]{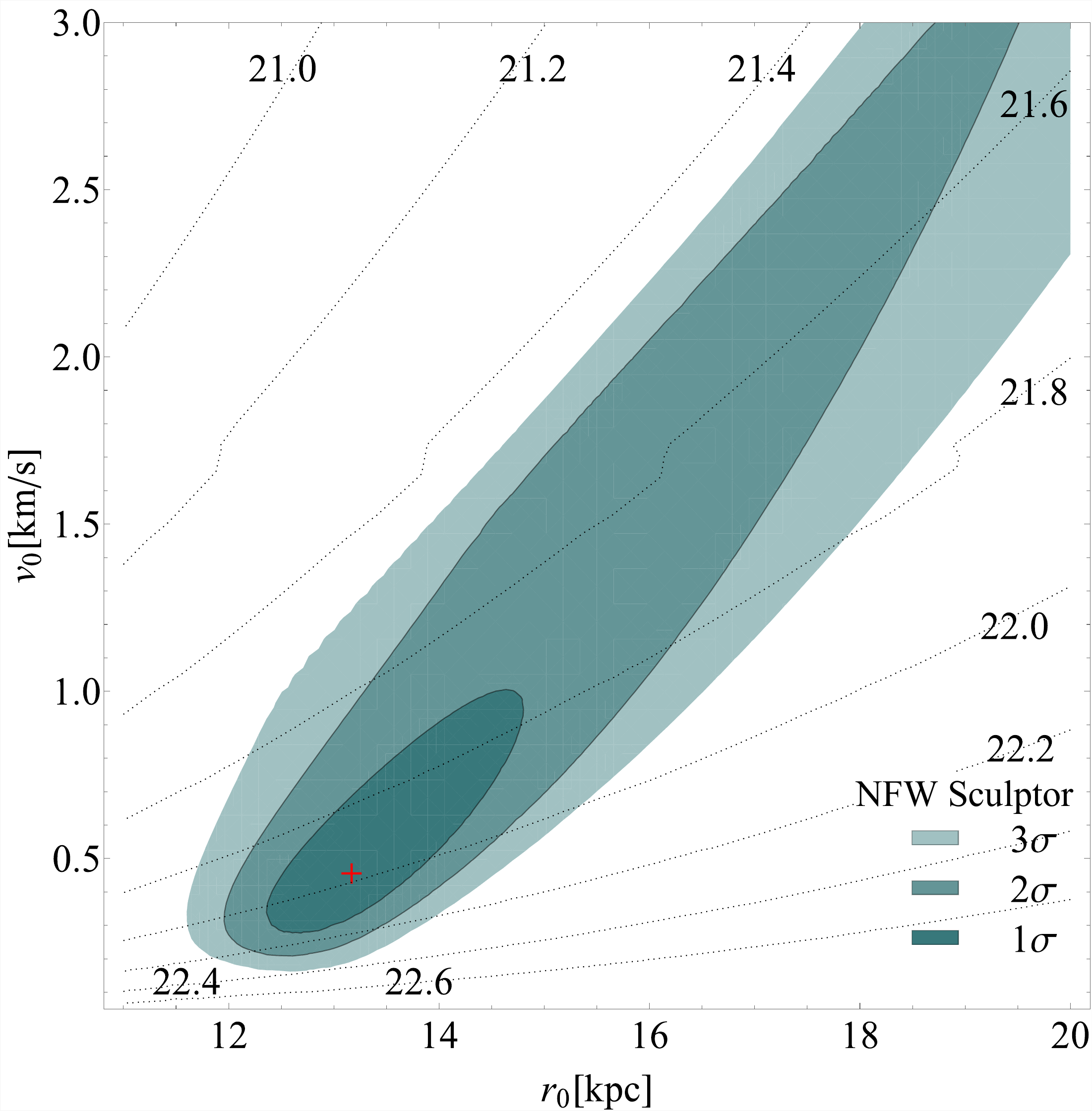}
\end{minipage}
\begin{minipage}[t]{0.49\linewidth}
\centering
\includegraphics[width=0.9\textwidth]{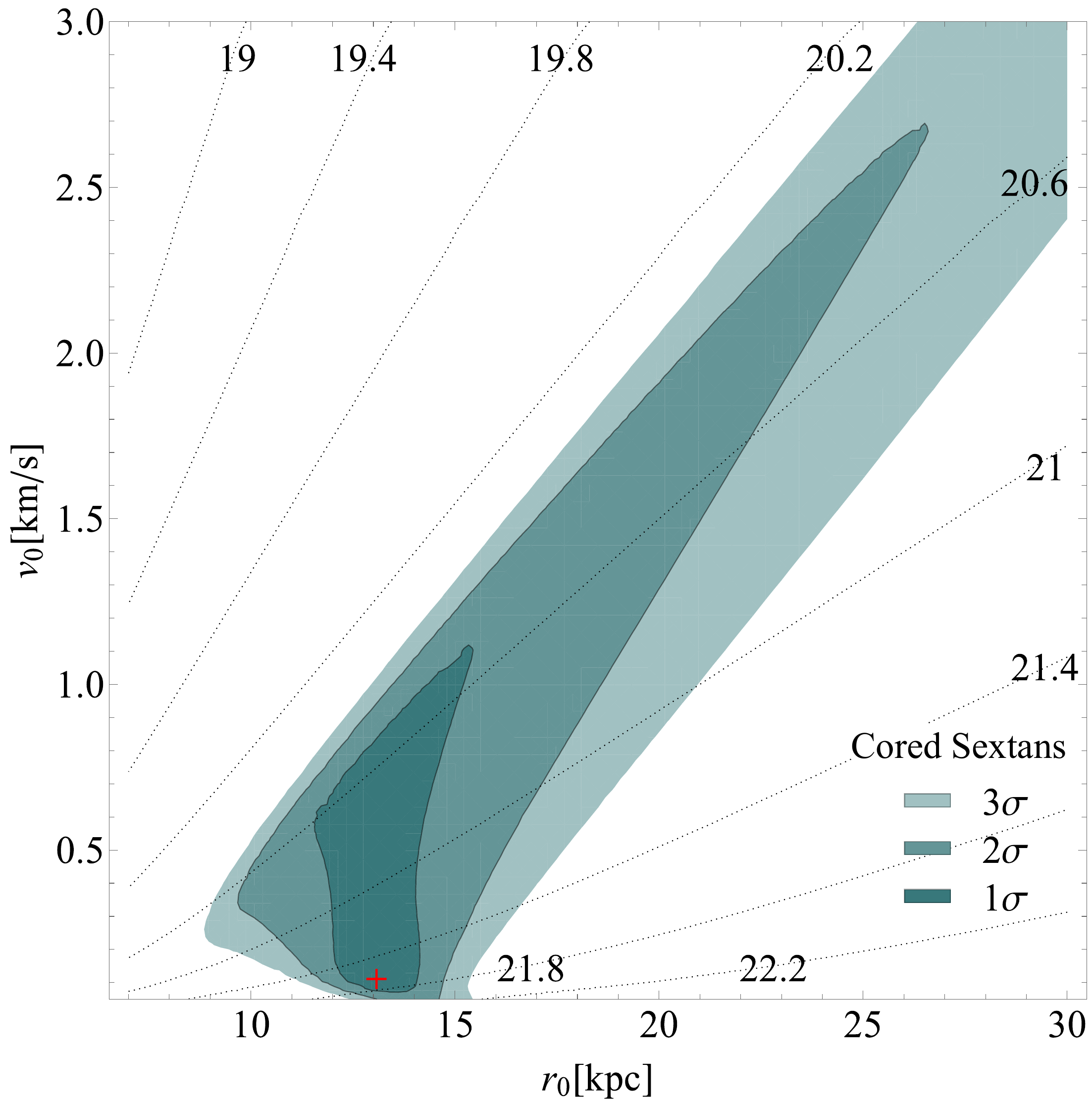}
\end{minipage}
\begin{minipage}[t]{0.49\linewidth}
\centering
\includegraphics[width=0.9\textwidth]{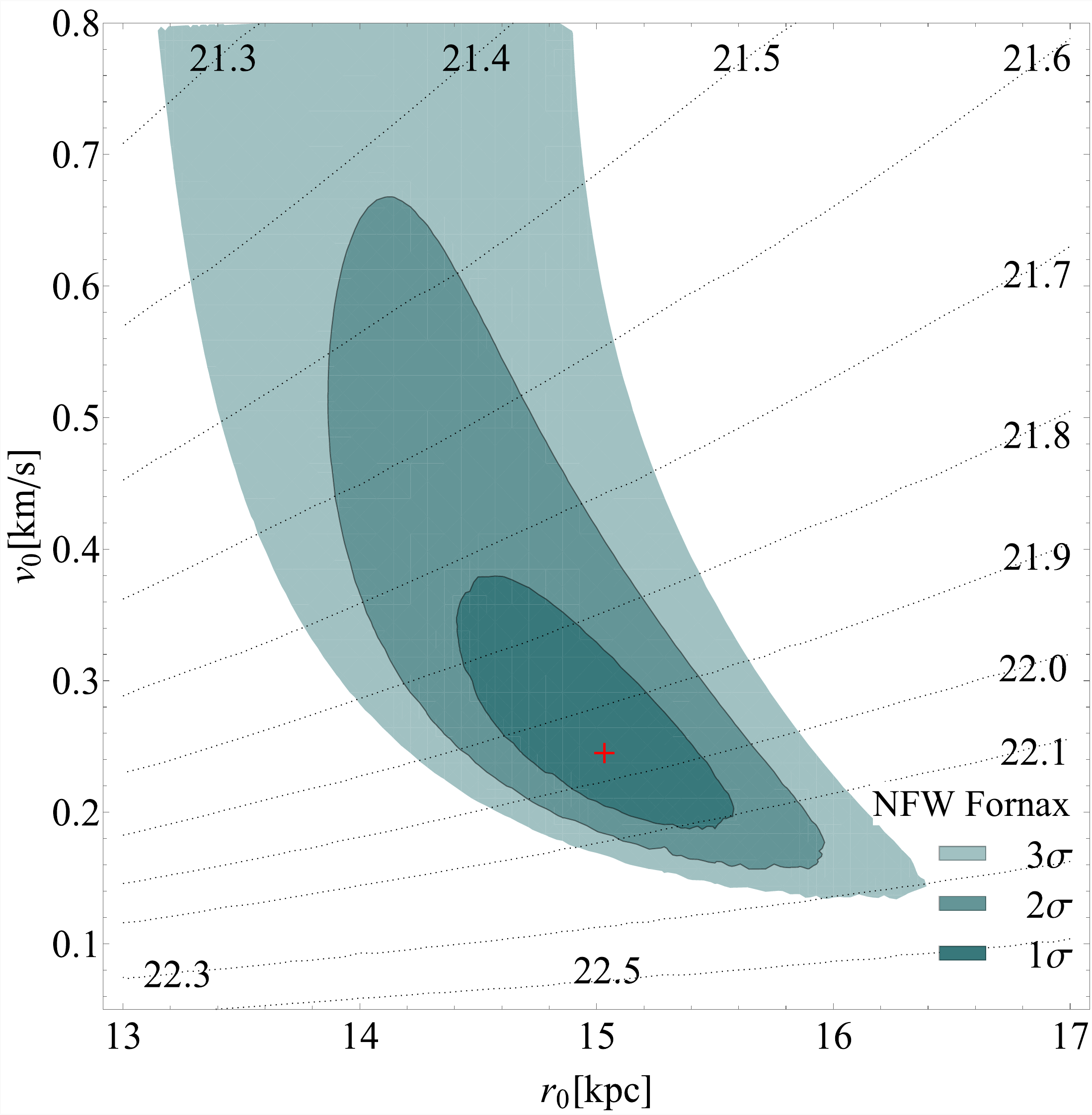}
\end{minipage}
\end{center}
\caption{1$\sigma$, $2\sigma$ and $3\sigma$ confidence intervals in the $(r_0,v_0)$ plane.~Top panels refer to the Carina (left) and Sculptor (right) dSphs, whereas bottom panels refer to the Sextans (left) and Fornax (right) dSphs.~In the case of the Sculptor and Fornax dSphs, we assume a NFW profile, whereas for the Carina and Sextans dSphs we assume a cored Zhao profile.~These four dSphs were chosen since they are characterised by a large number of stars for which kinematic data are available; see Tab.~\ref{tab:J1}.~In all cases we set $\epsilon_\phi=10^{-4}$.~Coloured contours are confidence intervals obtained from Eqs.~(\ref{eq:TS}) and (\ref{eq:chi}) with $\mathscr{L}_{\rm 1D}$ replaced by $\mathscr{L}_{2{\rm D}}$, and $\chi_1^2$ replaced by $\chi_2^2$, i.e.~the chi-squared distribution for 2 degrees of freedom.~In the four panels, a red cross represents the best fit point in the $(r_0,v_0)$ plane.}
\label{fig:2D}
\end{figure*}

\begin{figure*}[t]
\begin{center}
\begin{minipage}[t]{0.49\linewidth}
\centering
\includegraphics[width=\textwidth]{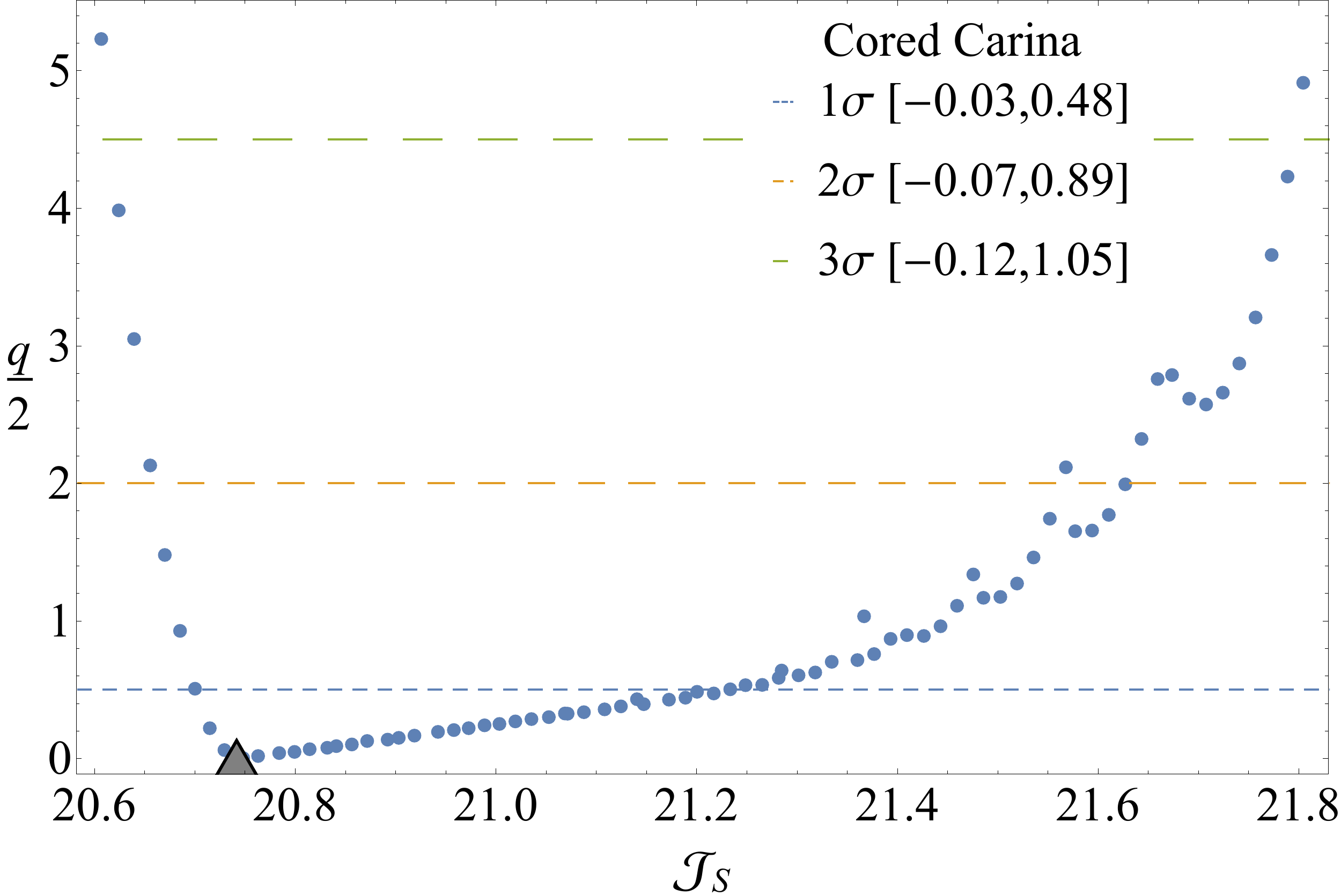}
\end{minipage}
\begin{minipage}[t]{0.49\linewidth}
\centering
\includegraphics[width=0.99\textwidth]{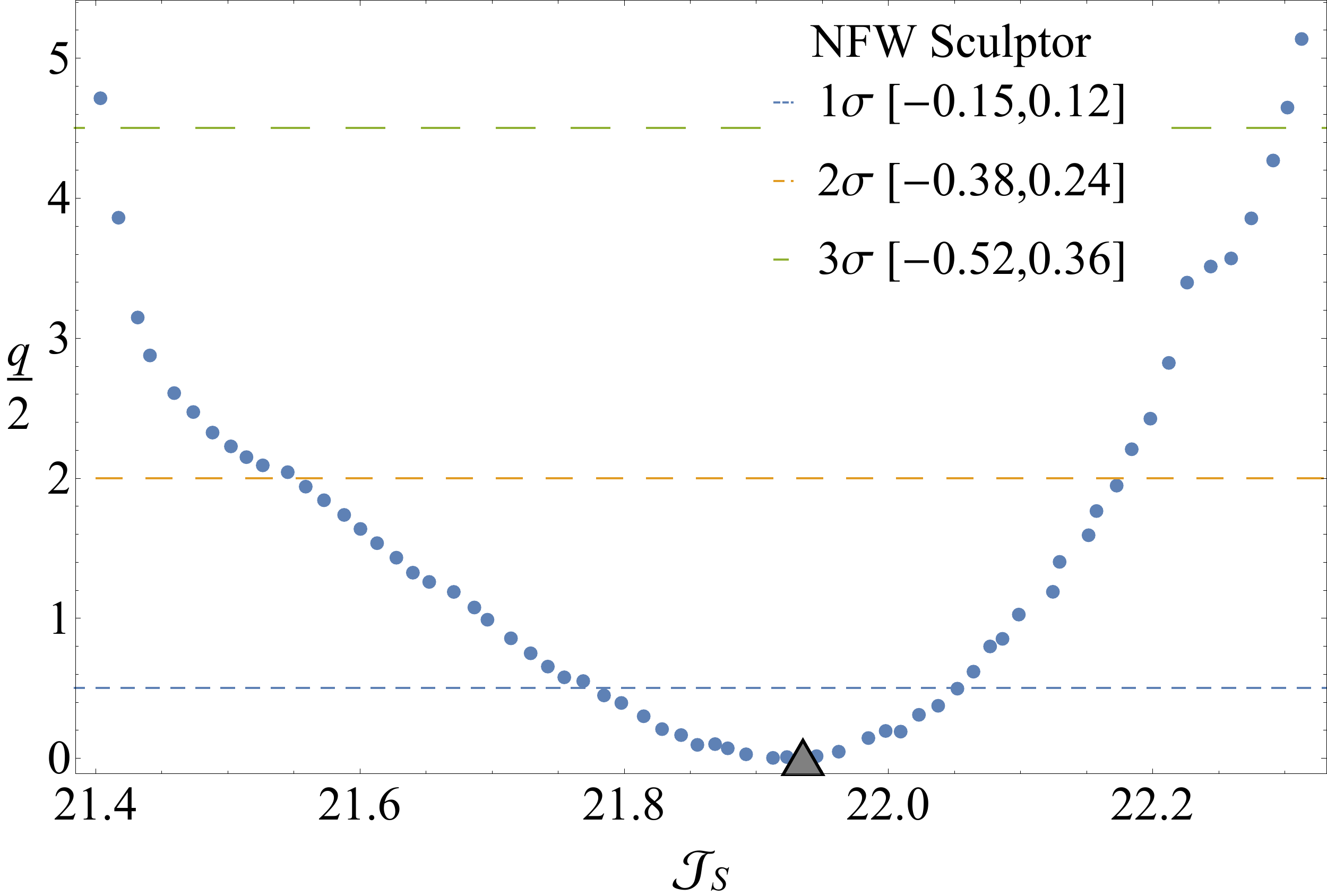}
\end{minipage}
\begin{minipage}[t]{0.49\linewidth}
\centering
\includegraphics[width=\textwidth]{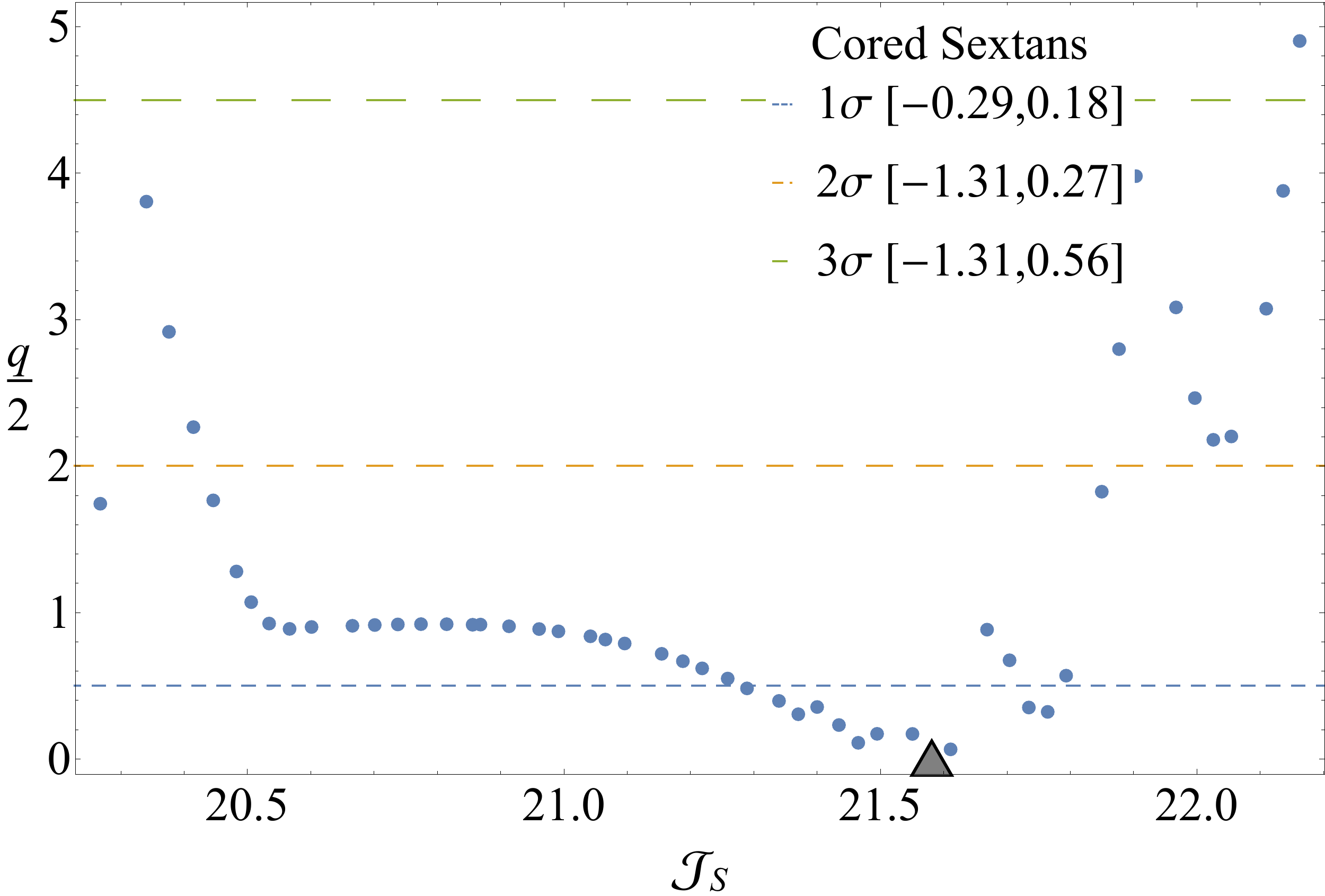}
\end{minipage}
\begin{minipage}[t]{0.49\linewidth}
\centering
\includegraphics[width=\textwidth]{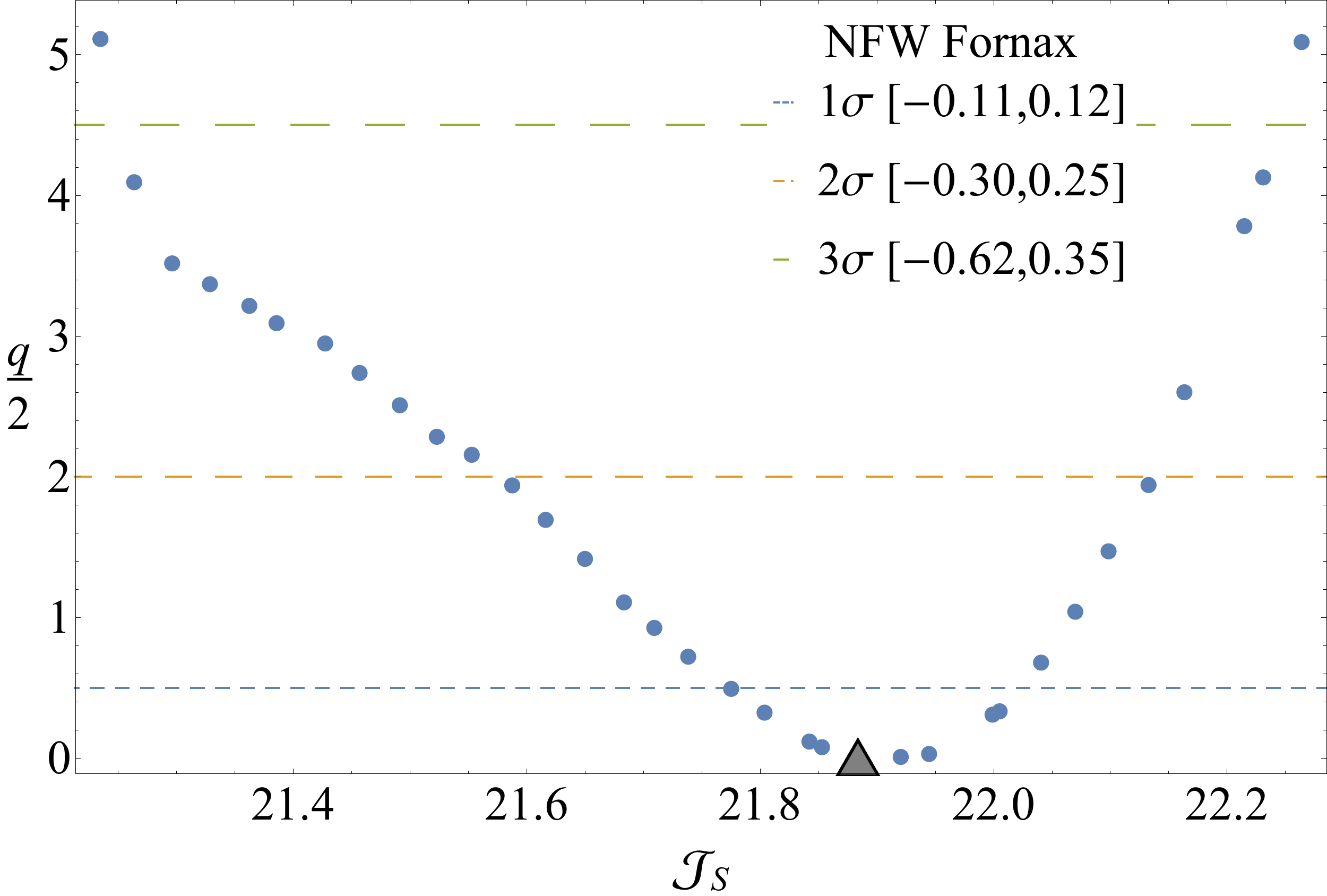}
\end{minipage}
\end{center}
\caption{Log-likelihood ratio, Eq.~(\ref{eq:TS}), as a function of $\mathcal{J}_S\equiv\log_{10}[J_S/({\rm GeV}^2 {\rm cm}^{-5})]$ for the Carina (top left), Sculptor (top right), Sextans (bottom left) and Fornax (bottom right) dSphs.~In the figure, the left (right) panels refer to a cored (NFW) DM profile.~Coloured dashed lines correspond to 1$\sigma$, $2\sigma$ and $3\sigma$ confidence intervals obtained from Eqs.~(\ref{eq:TS}) and (\ref{eq:chi}) as explained in Sec.~\ref{sec:analysis}.~Triangles represent the best fit points for $\mathcal{J}_S$.~In all cases, calculations are performed assuming $\alpha_\chi=10^{-2}$ and $\epsilon_\phi=10^{-4}$.\label{fig:1D}}  
\end{figure*}

\begin{figure*}[t]
\begin{center}
\begin{minipage}[t]{0.49\linewidth}
\centering
\includegraphics[width=\textwidth]{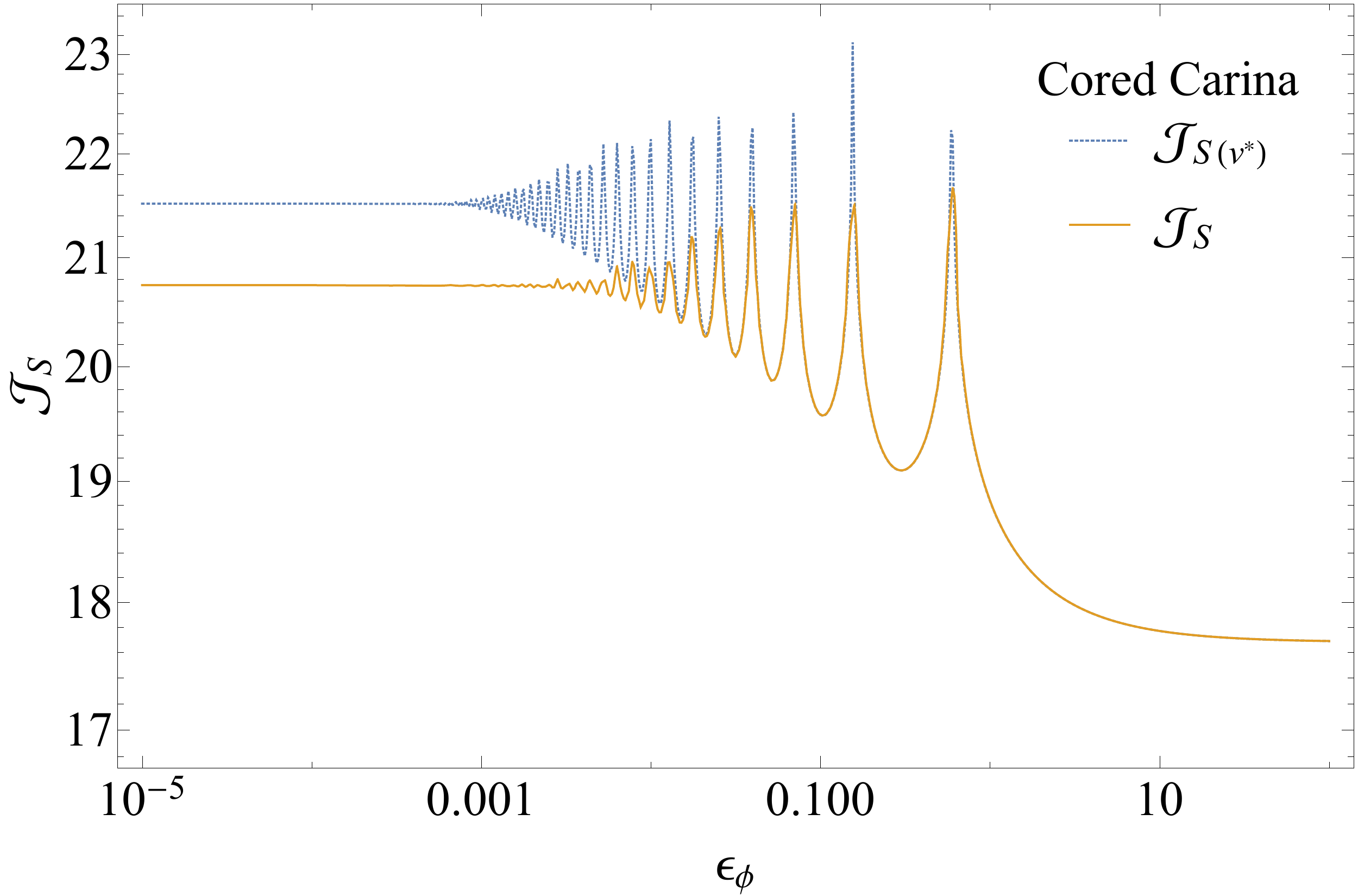}
\end{minipage}
\begin{minipage}[t]{0.49\linewidth}
\centering
\includegraphics[width=\textwidth]{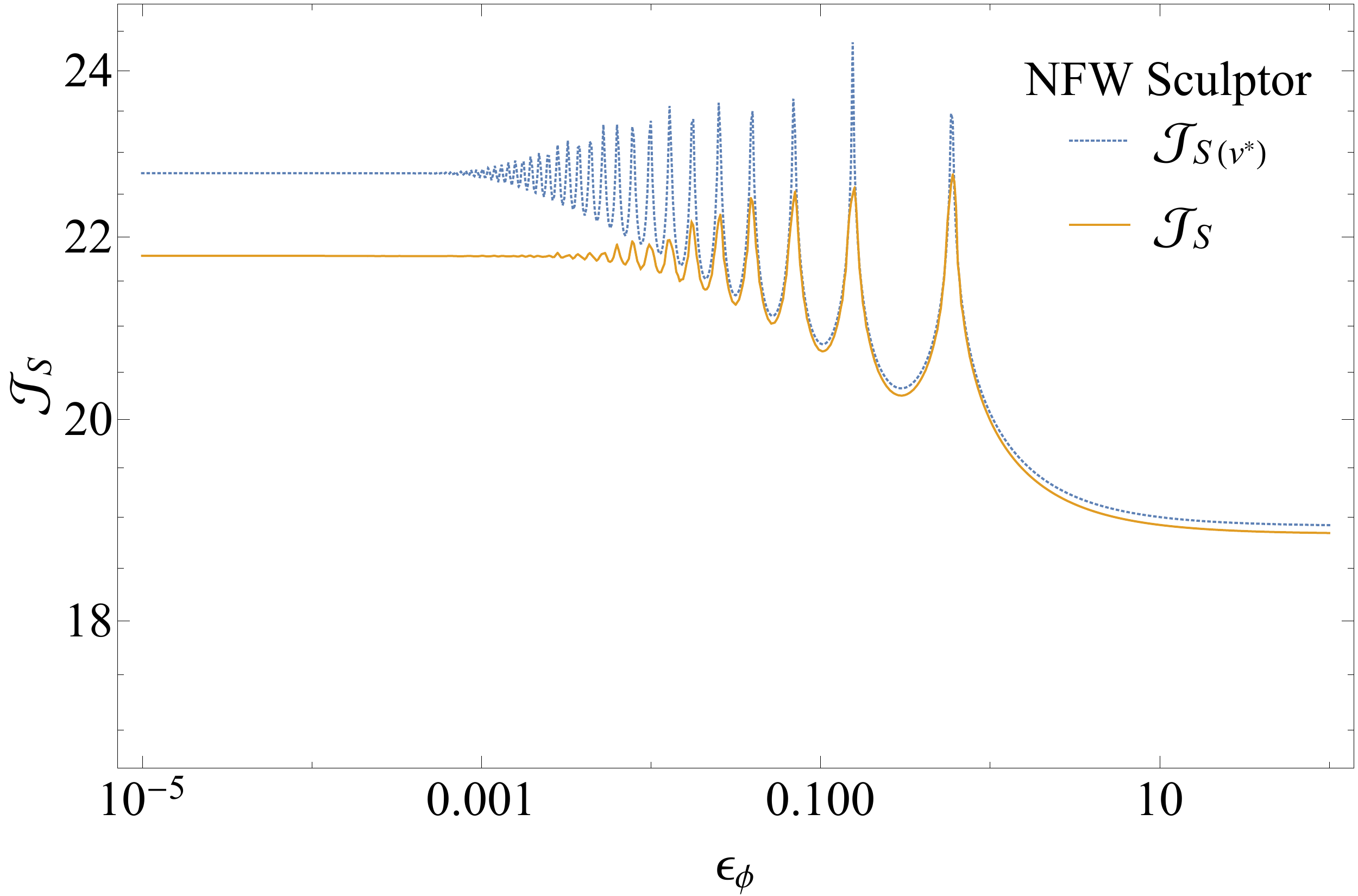}
\end{minipage}
\begin{minipage}[t]{0.49\linewidth}
\centering
\includegraphics[width=\textwidth]{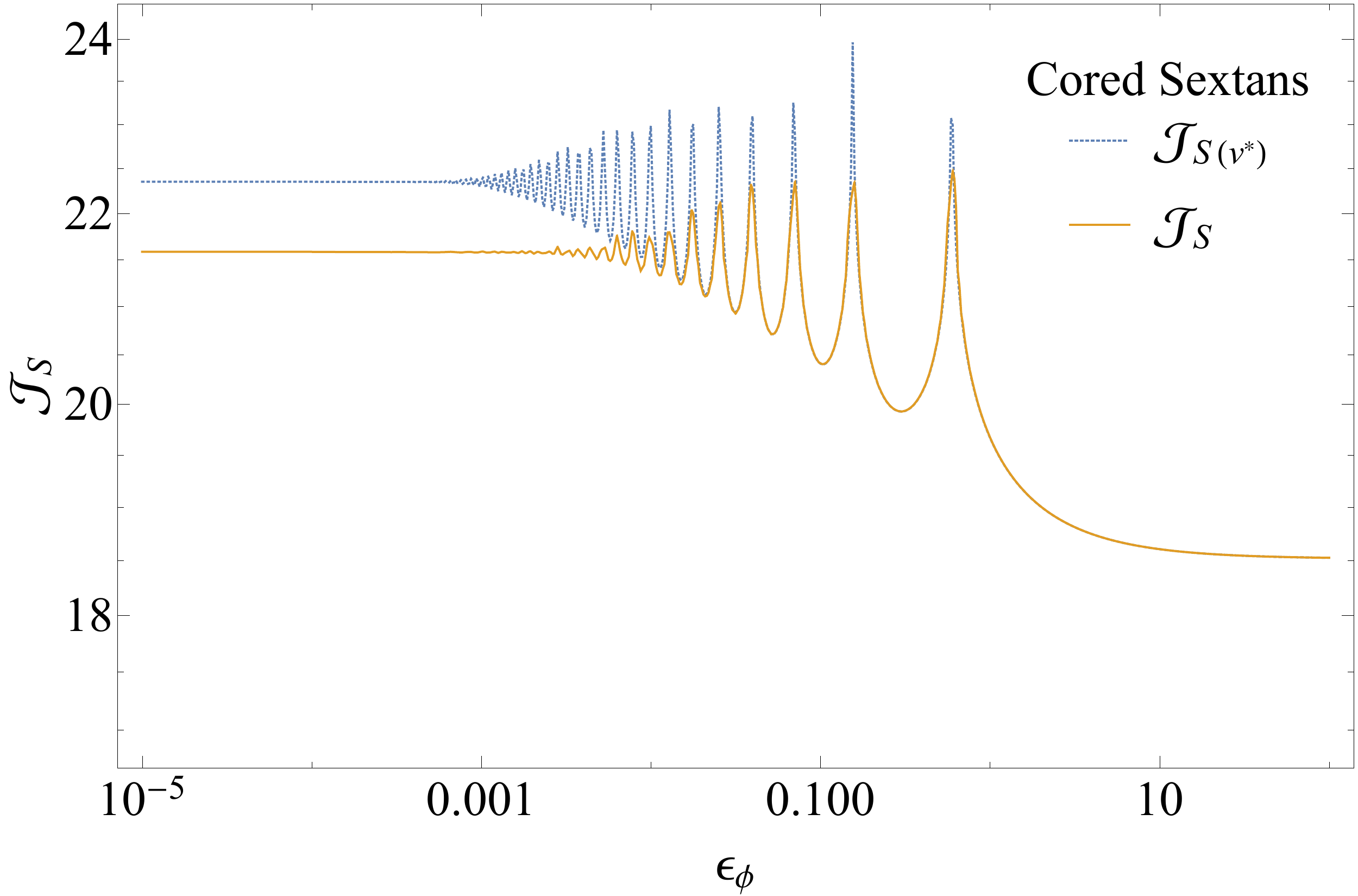}
\end{minipage}
\begin{minipage}[t]{0.49\linewidth}
\centering
\includegraphics[width=\textwidth]{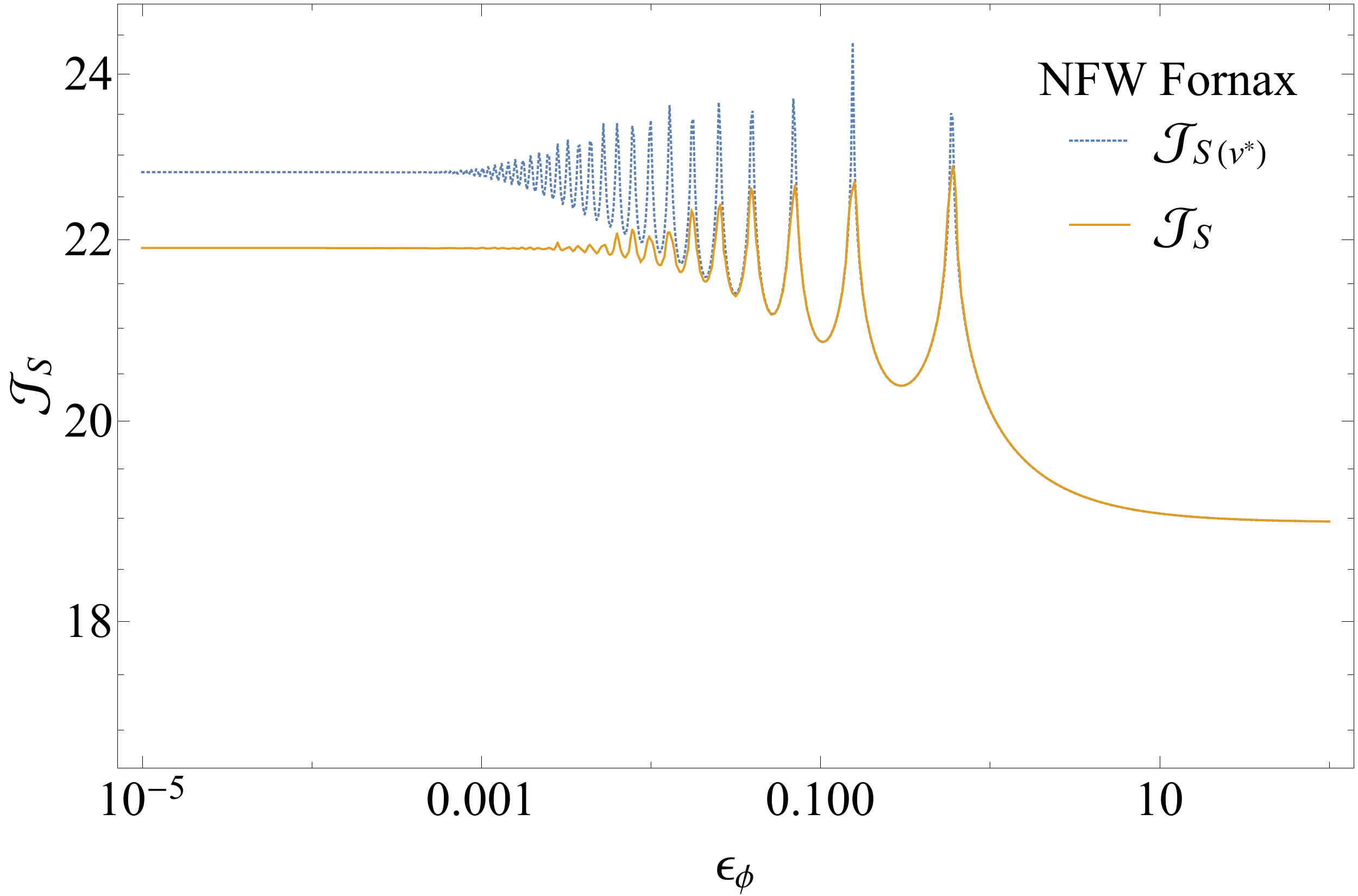}
\end{minipage}
\end{center}
\caption{$\mathcal{J}_S\equiv\log_{10}[J_S/({\rm GeV}^2 {\rm cm}^{-5})]$ as a function of $\epsilon_\phi$ ($\alpha_\chi=10^{-2}$) for the Carina (top left), Sculptor (top right), Sextans (bottom left) and Fornax (bottom right) dSphs.~The yellow line has been obtained using Eq.~(\ref{eq:Prel}) and a DM velocity distribution extracted from data as explained in Sec.~\ref{sec:DF}, the blue line corresponds to generalised $J$-factors computed under the approximation $S(v_{\rm rel})=S(v^*)$, $v^*=10^{-5}$ in natural units, according to which all DM particles in the dSph move with the same velocity.\label{fig:Jstar}}  
\end{figure*}

\section{Generalised J-factors from Stellar kinematic data}
\label{sec:analysis}

\subsection{Likelihood analysis}
Our method to determine the $J_S$ of a dSph from stellar kinematic data is based upon the likelihood function, $\mathscr{L}$~\cite{Chiappo:2016xfs}
\begin{equation}
-\ln\mathscr{L} = \frac{1}{2} \sum_{i=1}^{N_\star} \left[ \frac{(v_i-\bar{u})^2}{\sigma_i^2} +\ln\left( 2\pi\sigma_i^2 \right) \right]
\label{eq:L}
\end{equation}
where the index $i$ runs over the $N_\star$ stars in the dSph, $v_i$ is the line-of-sight velocity of the $i$-th star, and $\bar{u}$ is the systemic velocity of the galaxy; we approximate the latter with the mean stellar velocity of the sample.~A brief review of the used kinematic data is postponed to end of this section.~The expected velocity dispersion at the stellar projected distance to the galaxy centre $R_i$ is taken to be $\sigma_i^2=\epsilon_i^2+\sigma_{\rm los}^2(R_i)$, where $\epsilon_i$ is a measurement uncertainty, and, assuming isotropic stellar velocities, $\sigma_{\rm los}^2(R)$ reads~\cite{1987gady.book.....B}
\begin{equation}
\sigma_{\rm los}^2(R)=\frac{2G}{I(R)} \int_{R}^\infty {\rm d}r\,\frac{r}{\sqrt{r^2-R^2}} \int_r^\infty {\rm d}s \, \frac{\nu_\star(s) M(s)}{s^2}\,.
\end{equation}
In the above equation~\cite{1987gady.book.....B}
\begin{equation}
I(R)= 2\int_R^\infty {\rm d}r \frac{r}{\sqrt{r^2-R^2}} \,\nu_*(r)
\label{eq:plummer}
\end{equation}
is the surface brightness, and 
\begin{equation}
M(s) = 4\pi \int_0^s {\rm d}r\, r^2 \rho_\chi(r)
\end{equation}
is the DM mass enclosed in a sphere of radius $s$.~For the stellar density profile, $\nu_*$, we use Eq.~(\ref{eq:prof}) with $\rho_\chi\rightarrow\nu_\star$ and $(\alpha,\beta,\gamma)=(2,5,0)$ -- the so-called Plummer profile.~In this case, $r_\star$ and $\rho_\star$ denote scale radius and density, respectively.

The likelihood in Eq.~(\ref{eq:L}), $\mathscr{L}$, depends on the stellar kinematic data array ${\bm X}=({\bm x}_1,\dots,{\bm x}_{N_\star})$, where ${\bf x}_i=(R_i,v_i,\epsilon_i)$.~In principle, it also depends on four input parameters:~$\rho_0$ and $r_0$ for the DM component, and $\rho_\star$ and $r_\star$ for the stellar component.~However, $\rho_\star$ cancels in the $\nu_\star/I$ ratio in Eq.~(\ref{eq:L}), and will not be considered further.~In addition, the reference density $\rho_0$ will be replaced by the parameter $v_0=r_0\sqrt{G\rho_0}$.~Summarising, our likelihood function takes the following form: $\mathscr{L}=\mathscr{L}(v_0,r_0,r_\star|{\bm X})$.

For each dSphs in our sample, we derive $J_S$ and the associated statistical error through a profile likelihood approach.~The profile likelihood for $J_S$ is obtained from $\mathscr{L}$ numerically, as described in the following.~First, we construct a grid in the $(r_0,v_0)$ plane, and at each point $(r_0,v_0)$ of the grid maximise $\mathscr{L}$ over $r_\star$.~We denote by $\hat{r}_\star$ the point of maximum likelihood, and introduce the notation
\begin{equation}
\mathscr{L}_{2{\rm D}}(v_0,r_0|{\bm X})=\mathscr{L}(v_0,r_0,\hat{r}_\star|{\bm X})\,.
\end{equation}
At each point of the $(r_0,v_0)$ grid, we also calculate $J_S$.~Importantly, a degeneracy between $v_0$ and $r_0$ implies that different pairs of these parameters, and therefore different values of $\mathscr{L}_{2{\rm D}}$, can be associated with the same $J_S$.~Next, we divide the $J_S$-axis in bins.~At the central point of each bin, $J_S^c$, we associate the maximum value that $\mathscr{L}_{2{\rm D}}$ can have when $J_S$ varies in that bin.~Let us denote this maximum value of $\mathscr{L}_{2{\rm D}}$ by $\mathscr{L}_c$.~The function of $J_S^c$, $\mathscr{L}_{\rm 1D}(J_S^c|{\bm X})=\mathscr{L}_c$, is the discretised profile likelihood of $J_S$.~Through interpolation, we obtain the profile likelihood $\mathscr{L}_{\rm 1D}(J_S|{\bm X})$ at any $J_S$.~Analogously, $\mathscr{L}_{2{\rm D}}(v_0,r_0|{\bm X})$ represent the two-dimensional profile likelihood in the $(r_0,v_0)$ plane.

The value of $J_S$ maximising $\mathscr{L}_{\rm 1D}(J_S|{\bm X})$, $\hat{J}_S$, is our estimate for the generalised $J$-factor.~The error associated with $\hat{J}_S$ is computed numerically from the following test statistic
\begin{equation}
q(J_S)=-2\ln \frac{\mathscr{L}_{1{\rm D}}(J_S|{\bm X})}{\mathscr{L}_{1{\rm D}}(\hat{J}_{S}|{\bm X})}\,,
\label{eq:TS}
\end{equation}
which asymptotically obeys a $\chi^2_1$ distribution.~An $\alpha$\% confidence interval for $\hat{J}_{S}$ is then obtained by solving for $\Delta q$ the equation
\begin{equation}
\alpha = \int_{0}^{\Delta q} {\rm d}q\, \chi_1^2(q) \,,
\label{eq:chi}
\end{equation}
and imposing $q(J_S)\le \Delta q$.
The kinematic data used in this analysis were obtained through a series of surveys, entailing bolometric and spectroscopic measurements of the stellar population of dSphs (see \cite{2012AJ....144....4M, 2013pss5.book.1039W, Battaglia:2013wqa} and references therein for further information on dSphs kinematic data acquisition). The former observations produced information on the luminosity distribution of the system, which motivates the use of the Plummer profile. From the latter, the position of every star, together with the line-of-sight velocity and its uncertainty, are obtained. Finally, using the coordinates of the estimated centre of a dSph, the projected radial distance of every star can be evaluated. Combining all information results in the required data array $\bm{X}$.

\begin{table*}[t]
\centering
	\begin{ruledtabular}	
    		\begin{tabular}{llllllll}
    		\toprule
    		\boldmidrule
	 	\toprule
	  	\toprule
Galaxy & $N_\star$ & $\mathcal{J}$ (cored) &$\mathcal{J}_{S(v^*)}$ (cored) & $\mathcal{J}_{S}$ (cored) & $\mathcal{J}$ (NFW) & $\mathcal{J}_{S(v^*)}$ (NFW) & $\mathcal{J}_{S}$ (NFW)\\
\hline
Bootes I & 14 &$19.34_{-2.07}^{+0.38}$ & $23.17_{-2.07}^{+0.38}$ & $21.65_{-0.92}^{+0.34}$ & $17.95_{-0.74}^{+0.54}$ & $21.79_{-0.74}^{+0.54}$ & $21.13_{-0.48}^{+0.40}$\\
Leo IV & 17 &$16.46_{-0.61}^{+1.75}$ & $20.29_{-0.61}^{+1.75}$ & $19.89_{-0.45}^{+0.94}$ & $16.89_{-0.92}^{+0.83}$ & $20.73_{-0.92}^{+0.83}$ & $20.46_{-0.78}^{+0.65}$\\
Leo T & 19 &$17.45_{-0.95}^{+0.49}$ & $21.29_{-0.95}^{+0.49}$ & $20.53_{-0.84}^{+0.34}$ & $17.44_{-0.87}^{+0.43}$ & $21.28_{-0.87}^{+0.43}$ & $20.60_{-0.81}^{+0.37}$\\
Bootes II &  20 &$18.78_{-1.01}^{+1.46}$ & $22.61_{-1.01}^{+1.46}$ & $22.10_{-0.83}^{+1.00}$ & $18.89_{-1.11}^{+1.20}$ & $22.72_{-1.11}^{+1.20}$ & $22.21_{-0.89}^{+1.16}$\\
Ursa Major II & 20 & $20.29_{-0.72}^{+0.43}$ & $24.12_{-0.72}^{+0.43}$ & $22.77_{-0.28}^{+0.29}$ & $19.87_{-0.18}^{+0.27}$ & $23.71_{-0.18}^{+0.27}$ & $22.76_{-0.14}^{+0.25}$\\
Canes Venatici II & 25 &$18.53_{-0.74}^{+0.35}$ & $22.36_{-0.74}^{+0.35}$ & $21.23_{-0.50}^{+0.34}$ & $18.49_{-0.70}^{+0.31}$ & $22.32_{-0.70}^{+0.31}$ & $21.27_{-0.46}^{+0.23}$\\
Hercules & 30 &$18.00_{-0.29}^{+0.35}$ & $21.83_{-0.29}^{+0.35}$ & $21.14_{-0.21}^{+0.28}$ & $18.12_{-0.35}^{+0.27}$ & $21.95_{-0.35}^{+0.27}$ & $21.35_{-0.31}^{+0.22}$\\
Ursa Major I & 39 &$17.77_{-0.28}^{+0.80}$ & $21.60_{-0.28}^{+0.80}$ & $21.00_{-0.28}^{+0.59}$ & $18.22_{-0.58}^{+0.95}$ & $22.06_{-0.58}^{+0.95}$ & $21.52_{-0.70}^{+0.66}$\\
Willman 1 &45 & $19.40_{-0.45}^{+1.20}$ & $23.24_{-0.45}^{+1.20}$ & $22.43_{-0.24}^{+0.62}$ & $19.69_{-0.52}^{+0.31}$ & $23.52_{-0.52}^{+0.31}$ & $22.54_{-0.23}^{+0.29}$\\
Coma Berenices & 59 &$19.93_{-0.87}^{+0.77}$ & $23.77_{-0.87}^{+0.77}$ & $22.56_{-0.47}^{+0.36}$ & $19.42_{-0.45}^{+0.28}$ & $23.26_{-0.45}^{+0.28}$ & $22.35_{-0.31}^{+0.21}$\\
Segue 1 &66 & $19.10_{-0.30}^{+0.47}$ & $22.93_{-0.30}^{+0.47}$ & $22.39_{-0.23}^{+0.28}$ & $19.26_{-0.46}^{+0.48}$ & $23.09_{-0.46}^{+0.48}$ & $22.72_{-0.44}^{+0.42}$\\
Ursa Minor & 196 &$19.47_{-1.04}^{+0.22}$ & $23.31_{-1.04}^{+0.22}$ & $22.46_{-1.29}^{+0.18}$ & $19.57_{-0.25}^{+0.08}$ & $23.41_{-0.25}^{+0.08}$ & $22.62_{-0.27}^{+0.06}$\\
Canes Venatici I & 214 &$17.88_{-0.99}^{+0.19}$ & $21.72_{-0.99}^{+0.19}$ & $20.91_{-0.99}^{+0.19}$ & $18.01_{-0.29}^{+0.28}$ & $21.84_{-0.29}^{+0.28}$ & $21.11_{-0.25}^{+0.29}$\\
Leo I & 328 &$17.53_{-0.10}^{+0.22}$ & $21.36_{-0.10}^{+0.22}$ & $20.43_{-0.04}^{+0.25}$ & $17.68_{-0.17}^{+0.23}$ & $21.52_{-0.17}^{+0.23}$ & $20.56_{-0.13}^{+0.29}$\\
Draco & 353 &$18.59_{-0.13}^{+0.20}$ & $22.42_{-0.13}^{+0.20}$ & $21.36_{-0.03}^{+0.30}$ & $18.78_{-0.26}^{+0.21}$ & $22.61_{-0.26}^{+0.21}$ & $21.65_{-0.16}^{+0.23}$\\
Sextans &424 & $18.52_{-0.29}^{+0.19}$ & $22.35_{-0.29}^{+0.19}$ & $21.58_{-0.29}^{+0.18}$ & $18.73_{-0.19}^{+0.22}$ & $22.57_{-0.19}^{+0.22}$ & $21.86_{-0.18}^{+0.16}$\\
Carina &758 & $17.68_{-0.07}^{+0.44}$ & $21.51_{-0.07}^{+0.44}$ & $20.74_{-0.03}^{+0.48}$ & $17.71_{-0.02}^{+0.79}$ & $21.54_{-0.02}^{+0.79}$ & $20.84_{-0.02}^{+0.86}$\\
Sculptor & 1352 &$18.68_{-0.22}^{+0.14}$ & $22.52_{-0.22}^{+0.14}$ & $21.63_{-0.23}^{+0.15}$ & $18.92_{-0.14}^{+0.10}$ & $22.76_{-0.14}^{+0.10}$ & $21.94_{-0.15}^{+0.12}$\\
Sagittarius & 1373 &$19.77_{-0.17}^{+0.16}$ & $23.61_{-0.17}^{+0.16}$ & $22.51_{-0.16}^{+0.16}$ & $20.25_{-0.12}^{+0.09}$ & $24.09_{-0.12}^{+0.09}$ & $23.16_{-0.11}^{+0.09}$\\
Fornax & 2409 &$18.70_{-0.23}^{+0.13}$ & $22.54_{-0.23}^{+0.13}$ & $21.59_{-0.20}^{+0.11}$ & $18.94_{-0.07}^{+0.08}$ & $22.77_{-0.07}^{+0.08}$ & $21.88_{-0.11}^{+0.12}$\\
  		\bottomrule
      		\bottomrule
      		\boldmidrule
    		\end{tabular}
    	\end{ruledtabular}	
        \caption{Table reporting canonical and generalised $J$-factors computed for the particle physics input parameters $\epsilon_\phi=10^{-4}$ and $\alpha_\chi=10^{-2}$.~For canonical $J$-factors, $S=1$ and $\mathcal{J}_S=\mathcal{J}$.~This calculation has been performed for the 20 dSphs in the first column, and for both NFW and cored Zhao DM profiles.~Tables corresponding to different choices of particle physics inputs are provided with the online version of this article as a supplementary material.~Comparing generalised and canonical $J$-factors in the tables, we find significant differences -- up to several order of magnitudes in all cases e.g. three in the case of the Fornax dSph.~The table also shows the generalised $J$-factors computed under the approximation $S(v_{\rm rel})=S(v^*)$, with $v^{*}=10^{-5}$ in natural units.~In this case, $\mathcal{J}_S=\mathcal{J}_{S(v^*)}.~$We find that the $S(v_{\rm rel})=S(v^*)$ approximation overestimates $J_S$ by up to one order of magnitude.}
        \label{tab:J1}
\end{table*}

\begin{table*}[t]
\centering
	\begin{ruledtabular}	
    		\begin{tabular}{llllllll}
    		\toprule
    		\boldmidrule
	 	\toprule
	  	\toprule
Galaxy & $N_\star$ & $\mathcal{J}$ (cored) &$\mathcal{J}_{S(v^*)}$ (cored) & $\mathcal{J}_{S}$ (cored) & $\mathcal{J}$ (NFW) & $\mathcal{J}_{S(v^*)}$ (NFW) & $\mathcal{J}_{S}$ (NFW)\\
\hline
Bootes I & 14 &$19.34_{-2.07}^{+0.38}$ & $21.23_{-2.07}^{+0.38}$ & $21.19_{-2.03}^{+0.37}$ & $17.95_{-0.74}^{+0.54}$ & $19.85_{-0.74}^{+0.54}$ & $19.84_{-0.75}^{+0.54}$\\
Leo IV & 17 &$16.46_{-0.61}^{+1.75}$ & $18.35_{-0.61}^{+1.75}$ & $18.35_{-0.61}^{+1.74}$ & $16.89_{-0.92}^{+0.83}$ & $18.79_{-0.92}^{+0.83}$ & $18.79_{-0.92}^{+0.92}$\\
Leo T & 19 &$17.45_{-0.95}^{+0.49}$ & $19.35_{-0.95}^{+0.49}$ & $19.34_{-0.95}^{+0.49}$ & $17.44_{-0.87}^{+0.43}$ & $19.34_{-0.87}^{+0.43}$ & $19.34_{-0.87}^{+0.49}$\\
Bootes II & 20 &$18.78_{-1.01}^{+1.46}$ & $20.67_{-1.01}^{+1.46}$ & $20.67_{-1.01}^{+1.44}$ & $18.89_{-1.11}^{+1.20}$ & $20.78_{-1.11}^{+1.20}$ & $20.78_{-1.11}^{+1.19}$\\
Ursa Major II &20 & $20.29_{-0.72}^{+0.43}$ & $22.19_{-0.72}^{+0.43}$ & $22.17_{-0.78}^{+0.40}$ & $19.87_{-0.18}^{+0.27}$ & $21.77_{-0.18}^{+0.27}$ & $21.76_{-0.19}^{+0.26}$\\
Canes Venatici II & 25 &$18.53_{-0.74}^{+0.35}$ & $20.42_{-0.74}^{+0.35}$ & $20.42_{-0.73}^{+0.35}$ & $18.49_{-0.70}^{+0.31}$ & $20.38_{-0.70}^{+0.31}$ & $20.37_{-0.69}^{+0.31}$\\
Hercules & 30 &$18.00_{-0.29}^{+0.35}$ & $19.89_{-0.29}^{+0.35}$ & $19.89_{-0.29}^{+0.35}$ & $18.12_{-0.35}^{+0.27}$ & $20.01_{-0.35}^{+0.27}$ & $20.01_{-0.35}^{+0.26}$\\
Ursa Major I &39 & $17.77_{-0.28}^{+0.80}$ & $19.66_{-0.28}^{+0.80}$ & $19.66_{-0.28}^{+0.80}$ & $18.22_{-0.58}^{+0.95}$ & $20.12_{-0.58}^{+0.95}$ & $20.12_{-0.58}^{+0.95}$\\
Willman 1 & 45 &$19.40_{-0.45}^{+1.20}$ & $21.30_{-0.45}^{+1.20}$ & $21.29_{-0.42}^{+1.19}$ & $19.69_{-0.52}^{+0.31}$ & $21.59_{-0.52}^{+0.31}$ & $21.58_{-0.51}^{+0.30}$\\
Coma Berenices &  59 &$19.93_{-0.87}^{+0.77}$ & $21.83_{-0.87}^{+0.77}$ & $21.82_{-0.90}^{+0.74}$ & $19.42_{-0.45}^{+0.28}$ & $21.32_{-0.45}^{+0.28}$ & $21.31_{-0.48}^{+0.28}$\\
Segue 1 &  66 &$19.10_{-0.30}^{+0.47}$ & $20.99_{-0.30}^{+0.47}$ & $20.99_{-0.30}^{+0.46}$ & $19.26_{-0.46}^{+0.48}$ & $21.15_{-0.46}^{+0.48}$ & $21.15_{-0.46}^{+0.47}$\\
Ursa Minor & 196 &$19.47_{-1.04}^{+0.22}$ & $21.37_{-1.04}^{+0.22}$ & $21.37_{-1.05}^{+0.19}$ & $19.57_{-0.25}^{+0.08}$ & $21.47_{-0.25}^{+0.08}$ & $21.47_{-0.25}^{+0.08}$\\
Canes Venatici I & 214 &$17.88_{-0.99}^{+0.19}$ & $19.78_{-0.99}^{+0.19}$ & $19.77_{-0.99}^{+0.19}$ & $18.01_{-0.29}^{+0.28}$ & $19.90_{-0.29}^{+0.28}$ & $19.90_{-0.22}^{+0.28}$\\
Leo I & 328 & $17.53_{-0.10}^{+0.22}$ & $19.42_{-0.10}^{+0.22}$ & $19.42_{-0.10}^{+0.22}$ & $17.68_{-0.17}^{+0.23}$ & $19.58_{-0.17}^{+0.23}$ & $19.57_{-0.11}^{+0.22}$\\
Draco & 353 &$18.59_{-0.13}^{+0.20}$ & $20.48_{-0.13}^{+0.20}$ & $20.47_{-0.15}^{+0.23}$ & $18.78_{-0.26}^{+0.21}$ & $20.67_{-0.26}^{+0.20}$ & $20.66_{-0.26}^{+0.21}$\\
Sextans & 424 &$18.52_{-0.29}^{+0.19}$ & $20.41_{-0.29}^{+0.19}$ & $20.41_{-0.29}^{+0.19}$ & $18.73_{-0.19}^{+0.22}$ & $20.63_{-0.19}^{+0.22}$ & $20.63_{-0.19}^{+0.22}$\\
Carina & 758 &$17.68_{-0.07}^{+0.44}$ & $19.58_{-0.07}^{+0.44}$ & $19.57_{-0.05}^{+0.42}$ & $17.71_{-0.02}^{+0.79}$ & $19.60_{-0.02}^{+0.79}$ & $19.60_{-0.02}^{+0.83}$\\
Sculptor & 1352 &$18.68_{-0.22}^{+0.14}$ & $20.58_{-0.22}^{+0.14}$ & $20.58_{-0.22}^{+0.14}$ & $18.92_{-0.14}^{+0.10}$ & $20.82_{-0.14}^{+0.10}$ & $20.81_{-0.14}^{+0.10}$\\
Sagittarius & 1373 &$19.77_{-0.17}^{+0.16}$ & $21.67_{-0.17}^{+0.16}$ & $21.66_{-0.17}^{+0.16}$ & $20.25_{-0.12}^{+0.09}$ & $22.15_{-0.12}^{+0.09}$ & $22.14_{-0.11}^{+0.11}$\\
Fornax & 2409 &$18.70_{-0.23}^{+0.13}$ & $20.60_{-0.23}^{+0.13}$ & $20.59_{-0.23}^{+0.13}$ & $18.94_{-0.07}^{+0.08}$ & $20.83_{-0.07}^{+0.08}$ & $20.83_{-0.07}^{+0.08}$\\
  		\bottomrule
      		\bottomrule
      		\boldmidrule
    		\end{tabular}
    	\end{ruledtabular}	
        \caption{Same as for Tab.~\ref{tab:J1}, but now for $\epsilon_\phi=0.1$.}
        \label{tab:J2}
\end{table*}

\section{Results}
\label{sec:results}
\label{sec:results}
In this section we calculate the best fit values for $r_0$, $v_0$ and $J_S$ from the profile likelihoods $\mathscr{L}_{2{\rm D}}$ and $\mathscr{L}_{\rm 1D}$.~We perform this calculation for a sample of 20 dSphs (see, e.g.~Tab.~\ref{tab:J1}), and use methods and data described in the previous section.~Results are presented for selected values of the parameters $\alpha_\chi$, which determines $\epsilon_v$, and $\epsilon_\phi$.~The parameter $\alpha_\chi$ is set to the reference value $10^{-2}$, since in the $\alpha_\chi\ll 1$ limit corrections to $S$ due to DM bound state formation are only important at resonance, i.e.~for $\epsilon_\phi \simeq 6/(\pi^2 n^2)$, $n\in \mathbb{Z}^+$~\cite{Boddy:2017vpe}.~Values of $J_S$ corresponding to different choices of $\alpha_\chi$ can be obtained from the expressions reported at the end of Sec.~\ref{sec:theory}.~For the $\epsilon_\phi$ parameter, we focus on the range $[10^{-4},10^2]$.~For $\epsilon_\phi > 10^{-6} \alpha_\chi^{-1}$, constraints on the DM annihilation cross-section from CMB data~\cite{Galli:2011rz} are not strengthened by $S\neq 1$~\cite{Boddy:2017vpe}.

Fig.~\ref{fig:2D} shows 1$\sigma$, $2\sigma$ and $3\sigma$ confidence intervals in the $(r_0,v_0)$ plane.~Top panels refer to the Carina (left) and Sculptor (right) dSphs, whereas bottom panels refer to the Sextans (left) and Fornax (right) dSphs.~These four galaxies were chosen since they have a large $N_\star$ (see Tab.~\ref{tab:J1} and Tab.~\ref{tab:J2}).~In the case of the Sculptor and Fornax dSphs, we assume a NFW profile, whereas for the Carina and Sextans dSphs we consider a cored profile.~In all cases we set $\epsilon_\phi$ to the reference value $\epsilon_\phi=10^{-4}$.~Confidence intervals are obtained from Eqs.~(\ref{eq:TS}) and (\ref{eq:chi}) with $\mathscr{L}_{\rm 1D}$ replaced by $\mathscr{L}_{2{\rm D}}$, and $\chi_1^2$ replaced by $\chi_2^2$, where $\chi_2^2$ is the chi-squared distribution for 2 degrees of freedom.~In all panels, a red cross represents the best fit point, whereas coloured contours correspond to the associated two-dimensional confidence intervals.~While in the case of the Fornax dSph data can constrain $r_0$ and $v_0$ effectively, in the case of, e.g., the Carina dSph confidence intervals cover a wide range of values for $r_0$.~Furthermore, the best fit values that we find for $r_0$ would in some cases be excluded by numerical N-body simulations (in particular in the case of NFW profiles)~\cite{Boddy:2017vpe}.~However, in this study we pursue a data driven approach, and therefore do not impose constraints on $r_0$ and $v_0$ from N-body simulations. 

Fig.~\ref{fig:1D} shows the log-likelihood ratio, Eq.~(\ref{eq:TS}), as a function of $\mathcal{J}_S\equiv\log_{10}[J_S/({\rm GeV}^2 {\rm cm}^{-5})]$ for the Carina (top left), Sculptor (top right), Sextans (bottom left) and Fornax (bottom right) dSphs.~In the figure, the left (right) panels refer to a NFW (cored) DM profile.~Coloured dashed lines correspond to 1$\sigma$, $2\sigma$ and $3\sigma$ confidence intervals obtained from Eqs.~(\ref{eq:TS}) and (\ref{eq:chi}) as explained in the previous section.~For some of the found likelihoods the obtained confidence intervals might not correspond exactly to the number of standard deviations reported in the legends, since the true distribution of $q$ could differ from a $\chi_1^2$ distribution.~Finally, the triangles in the four panels represent the best fit points for $\mathcal{J}_S$.~Best fit values for $\mathcal{J}_S$, for all dSphs considered here and for selected values of the $\alpha_\chi$ and $\epsilon_\phi$ parameters, are reported in Tabs.~\ref{tab:J1} and \ref{tab:J2}.~Additional tables corresponding to different choices of $\epsilon_\phi$ are provided with the online version of this article as a supplementary material.~In the tables we also include our estimates for the canonical $J$-factors.~Comparing generalised and canonical $J$-factors, we find significant differences -- up to several order of magnitudes in all cases e.g. three in the case of the Fornax dSph.

We conclude this section by comparing our calculations with the results obtained assuming $S(v_{\rm rel})=S(v^*)$, where $v^*$ is a reference velocity for the DM particles in the dSph, e.g.~$10^{-5}$ in natural units.~This is a common approximation in the study of self-interacting DM.~Fig.~\ref{fig:Jstar} shows $\mathcal{J}_S$ as a function of $\epsilon_\phi$ ($\alpha_\chi=10^{-2}$) for the Carina (top left), Sculptor (top right), Sextans (bottom left) and Fornax (bottom right) dSphs.~The yellow line has been obtained using Eq.~(\ref{eq:Prel}) and a DM velocity distribution extracted from data, the blue line corresponds to the approximation $S(v_{\rm rel})=S(v^*)$, according to which all DM particles in the dSph move with the same velocity.~We find that the $S(v_{\rm rel})=S(v^*)$ approximation overestimates $J_S$ by up to one order of magnitude for small $\epsilon_\phi$.~This result highlights the importance of computing $J_S$ properly accounting for the DM velocity distribution in dSphs.

\section{Conclusion}
\label{sec:conclusions}
We derived the generalised $J$-factor, $J_S$, of 20 dSphs from stellar kinematic data in the case of self-interacting DM.~We focused on a family of DM self-interactions described by a Yukawa potential in the non-relativistic limit.~We determined $J_S$ and associated statistical error within the profile likelihood approach proposed in~\cite{Chiappo:2016xfs}.~We performed our calculations for NFW and cored DM profiles, and for different combinations of particle physics input parameters, i.e.~$\alpha_\chi$ and $\epsilon_\phi$.~We found that canonical and generalised $J$-factors differ by up to several orders of magnitude for all dSphs considered in this study.~We also compared our results with a common approximation made when calculating $\gamma$-ray fluxes from dSphs, according to which all DM particles in dSphs move with the same velocity.~We found that this approximation overestimates $J_S$, with errors as large as one order of magnitude.~This study shows that a detailed model for the DM velocity distribution in dSphs is crucial in the calculation of $J_S$, and therefore in the experimental analysis and theoretical interpretation of $\gamma$-ray searches in dSphs.

\acknowledgments This work was supported by the Knut and Alice Wallenberg Foundation and is partly performed within the Swedish Consortium for Dark Matter Direct Detection (SweDCube).

%

\end{document}